\documentclass[]{aa}
\usepackage{psfig,amssymb}
\begin{document} 

\thesaurus{08.22.3 ; 08.16.4 ; 08.15.1 ; 08.06.3} 

\title{Period--Luminosity--Colour distribution and classification 
       of Galactic oxygen--rich LPVs\thanks{Based on data from the 
                                 HIPPARCOS astrometry satellite.}} 

\subtitle{II. Confrontation with pulsation models} 
\author{D. Barth\`es \and X. Luri} 
%
%
\institute{Departament d'Astronomia i Meteorologia, Universitat de Barcelona, 
          Avinguda Diagonal 647,  E--08028 Barcelona, Spain} 
\date { Received 28 January 2000 / Accepted ... 2000} 
\maketitle
\markboth{D. Barth\`es \& X. Luri: PLC distribution and classification of 
Galactic O--rich LPVs (II)}{}
\begin{abstract}

The kinematic and Period--Luminosity--Colour distribution of O-rich 
Long-Period Variable stars of the solar neighbourhood is interpreted in terms 
of pulsation modes, masses and metallicities. It is first shown that, because 
of input physics imperfections, the periods and mean colours derived from  the
existing linear and nonlinear nonadiabatic models must significantly  depart
from the actual behaviour of the stars. As a consequence systematic 
corrections have to be applied, as a first approximation, to our linear  model
grid. These free parameters, as well as the mixing length, are  calibrated on
the LPVs of the LMC and of some globular clusters, assuming  a mean mass of
$1\,M_\odot$ for the LMC Mira-like stars. Then, the  masses and metallicities
corresponding to the four kinematic/photometric  populations of local LPVs are
evaluated. The possibility of a varying   mixing-length parameter is discussed
and taken into account. Stars of the old  disk appear pulsating in the
fundamental mode\,: one group, mainly composed  of Miras, has mean mass
$<M>\;\simeq 0.9\,M_\odot$ and mean metallicity  $<Z>\;\simeq 0.02$, both
strongly increasing with the period\,; a second  group, slightly older and
mainly composed of SRb's, has  $<M>\;\simeq 0.9\,M_\odot$ and $<Z>\;\ga 
0.03$\,. Stars of the  thin disk appear pulsating in the first and second 
overtones, with  $<M_{1ov}>\;\approx 1.05\,M_\odot$, 
$<M_{2ov}>\;>0.75\,M_\odot$ and  $<Z>\;\ge 0.04$\,. Stars of the extended 
disk/halo appear
pulsating in  the fundamental mode, with $<M>\;\simeq 1.1\,M_\odot$ and 
$<Z>\;\simeq 0.01$. The mixing-length parameter probably decreases along the 
AGB by no more than 15\,\% per magnitude. The large, positive period 
corrections (more than 30\,\% for the fundamental and 8\,\% for the  first
overtone) that have to be applied to the LNA models used in this  study do not
seem to be explained by imperfect sub-photospheric physics alone,  especially
when nonlinear effects are taken into account. The origin of the  extra period
increase (at least 15\,\% for the fundamental mode) may be the  stellar wind,
which was neglected by all pulsation codes up to now. 

\end{abstract}

\section{Introduction} 

Among the most important and still unsolved issues concerning Long-Period 
Variable stars (LPV) are the modelling of their pulsation and even the mere 
identity of the predominant pulsation mode. Linear nonadiabatic (LNA) models 
are currently used on a wide scale for reasons of simplicity. Relying on 
LNA relations between the period and the fundamental parameters, and on 
dynamical models of the photospheric region, the pulsation mode of Mira 
stars is the fundamental according to the observed velocity amplitude 
(Hill \& Willson 1979, Willson 1982, Bowen 1988, Wood 1990), and the first 
overtone according to the angular diameter estimates (Haniff et al. 1995, 
Van Leeuwen et al. 1997), pulsation accelerations (Tuchman 1991) and the PL 
distribution of globular cluster stars (Feast 1996). One study based on the 
periods and luminosities of LPVs of the Large Magellanic Cloud (LMC) 
concluded that Miras were pulsating in the fundamental mode and low-amplitude 
semi-regulars on the first or second overtone (Wood \& Sebo 1996). However, 
another study, also taking into account a few Miras of the solar 
neighbourhood and various temperature estimates, supported the first overtone 
in a majority of Miras (Barth\`es 1998). Last, still relying on LNA models, 
Barth\`es \& Tuchman (1994) and Barth\`es \& Mattei (1997) found that the 
Fourier spectra of a few Miras were more easily explained than were the stars 
pulsating on the first overtone with the fundamental mode and other 
overtones also acting. 

In fact, the large pulsation amplitude and the complexity of the chemistry 
and radiative transfer in the photospheric region make the mean effective 
temperature and radius very uncertain, even for nearby stars (Bessell et al. 
1989a, 1996\,; Hofmann et al. 1998\,; Ya'ari \& Tuchman 1998). The 
{\sc Hipparcos} parallaxes, too, usually have large error bars, and they 
concern only about two dozen Miras. 

Last, most of these studies suffered from 
a strong sampling bias, favouring periods longer than 250 days, especially 
those ranging from 300 to 400 days.\\ 

On the other hand, hydrodynamical models predict that, after thermal 
relaxation, Miras pulsate in the fundamental mode with a period either much 
shorter than (Ya'ari \& Tuchman 1996, 1999) or very close to (Wood 1995, 
Bessell et al. 1996, Hofmann et al. 1998) the LNA period. These models mainly 
differ by their handling of the convective energy transport and of the 
equation of state, but they share the same unrealistic assumption\,: no wind
and  no extended circumstellar envelope at the outer boundary. As a
consequence,  the reliability of both the nonlinear and linear nonadiabatic
pulsation  models is uncertain, and the pulsation mode is still ambiguous.\\ 

An unfortunate result of these theoretical difficulties is that the present 
masses and metallicities of the LPVs have always been very uncertain. 
Indeed, no direct estimate of mass in binary systems has been possible up to 
now, and the chemistry and radiative transfer in the photospheric region are 
so complex that no reliable metallicity has been derived.\\ 

This paper mainly consists in exploiting the classification (based on 
kinematic and photometric criteria) and the period--luminosity--colour (PLC) 
distributions of the Oxygen-rich LPVs of the solar neighbourhood, which were 
determined in Paper~I of this series (Barth\`es et al. 1999). Its aims are 
to assess the pulsation models, and to identify the predominant mode and 
estimate the average mass and metallicity for each kinematic/photometric 
group. 

The data sets are presented in Sect.~2. They include the results of 
Paper~I, i.e. the periods, absorption-corrected absolute magnitudes and 
de-reddened colours of a sample of stars observed by {\sc Hipparcos}, but also 
the de-biased distributions of the four kinematic/photometric groups to which 
they belong. LPVs found in the LMC and in various globular clusters are also 
included, with a view to calibrating the free parameters of the models. 

Sect. 3 describes the pulsation models to which these data will be 
confronted\,: the linear nonadiabatic modelling code and the adopted 
temperature scale, i.e. colour--temperature--metallicity (CTZ) relations. 
Sect. 4 then explains the difficulties to be expected because of the 
physical approximations of the models, and also because of the nonlinear 
shape of the CTZ relations. We will be led to introduce some systematic 
correction parameters for the periods and colours. 

These free parameters, together with the mixing length, are calibrated in 
Sect. 5 by confronting the models with the PLC distributions of the LPVs in 
the LMC and globular clusters. The so-calibrated models are then confronted
with the PLC distribution of local stars in Sect. 6, and the pulsation mode,
mass  and metallicity are derived for each kinematic/photometric group.
Finally,  the stability and reliability of the results, as well as their
consequences  concerning the existing pulsation codes, are assessed in
Sect.~7. 

\section{Data} 

\subsection{Solar neighbourhood} 

Paper I has provided us with the mean absolute $K_0$ magnitudes and  $(V-K)_0$
indices of about 350 M-type LPVs of the solar neighbourhood.  For about 250
stars, $(J-K)_0$ is available. The $K$ magnitude is known  to closely mimic the
behaviour of bolometric magnitude. Moreover, as  mentioned in Paper I, simple
simulations of light curves have shown that the  mean $K$ magnitude and the
magnitude corresponding to the mean $K$ flux  agree with one another to within
a few percent (0.06 mag for a full amplitude  of 1 mag). The adopted magnitude
is thus a good representative of the actual  mean luminosity of each star. We
must recall however, that, because of the  paucity of the data, which were
taken at arbitrary phases, the observational  error on $K$ is
$\sigma=$0.25--0.50\,mag. 

The mean $V-K$ index was defined in Paper I as the difference between the 
midpoint $V$ magnitude and the mean $K$ magnitude. Simulations have shown 
that the midpoint value of $V$ is systematically larger than the mean by, 
usually, only a few hundredths of mag, or at the very most a few tenths. 
On the other hand, the magnitudes at minimum brightness tend to be 
underestimated by visual observers. However, this systematic error 
(a few tenths of mag\,; the half for the mean) depends altogether on the 
distance, the mean absolute magnitude and the amplitude (which is correlated 
to the brightness, but with an important scatter). Thus, globally over the 
sample of stars and the subsamples defined in Paper I, the systematic error 
on $<V>$ should not exceed $-0.1$ mag and thus may be neglected. 
Besides, the random error is about 0.3 mag. Summarizing, the $V-K$ data used 
in Paper~I are subject to a random error of about 0.5 mag. 

The luminosity calibrations performed in that paper led us to identify four 
groups, differing by their kinematics and/or luminosity\,: 
\begin{description}
 \item[Group 1:] mainly composed of Mira stars (84\%), with a kinematics 
                 corresponding to old disk stars\,; 

 \item[Group 2:] mainly composed of SRb stars (56\%), also with old disk 
                 kinematics\,; 

 \item[Group 3:] mainly composed by SRb stars (82\%), with a younger 
                 kinematics\,; 

 \item[Group 4:] a small group (14 stars, including 13 Miras) with a 
                 kinematics corresponding to extended-disk or halo stars. 
\end{description} 
The calibration procedure also provided us with the de-biased distributions 
of the populations corresponding to these groups. As explained in Paper~I, 
the overall error bars of their parameters were computed by means of many 
Monte-Carlo simulations, taking into account the above-mentioned photometric 
errors, the astrometric ones and those of the radial velocities. 
Concerning the barycenters of these distributions (i.e. the population 
means), the uncertainties are\,: 
about 0.1 mag (Groups 1, 2 \& 3) or 0.25 (Group 4) for $(V-K)_0$\,; about 
0.05 (Groups 1 \& 2), 0.08 (Group 3) or 0.1 (Group 4) for $M_K$.\\ 

Last, it must also be mentioned that, for a minor but significant proportion 
of semi-regular stars, the periods may by erroneous by as much as a factor of
2,  either because there are few data and/or we could not check them (see 
Paper~I), or because the light curves and/or Fourier spectra are ambiguous 
(two or three large-amplitude pseudo-periodicities liable to correspond to a 
mode -- see Mattei et al. [1997]). This means that the period scattering in 
the groups including a large proportion of semi-regulars (i.e. Groups 2 and  3)
is probably overestimated. Here again, the population mean should remain 
nearly unaffected.

\subsection{Globular cluster stars} 

For oxygen-rich LPV stars belonging to globular clusters (GC) of our Galaxy, 
periods, mean absolute bolometric magnitudes (derived from blackbody fits to 
mean, de-reddened $JHKL$ data) and mean $(J-K)_0$ index values have been 
found in Whitelock (1986) and Feast (1996). The $<m_{\rm bol}>$ 
precision is about 0.1--0.3 mag according to the amplitude and data sampling. 
That for $J-K$ is 0.03--0.15 as explained above, but most often 
$\simeq 0.05$. 

\subsection{LMC stars} 

As hundreds of O-rich Mira-like stars (i.e. LPVs with $\Delta I \le 0.9$ mag) 
have been observed in the Large Magellanic Cloud, we will handle them in a 
synthetic way, by considering their mean Period-Luminosity ($M_{\rm bol}$) 
and Period-Colour ($J-K$) relations, the $1\sigma$ scattering about it, and 
the barycenter of this population. The latter is defined by the mean period, 
already computed by Reid et al. (1995) and the corresponding mean bolometric 
magnitude and colour. 
The PL and PC relations were taken from Feast et al. (1989) and Hughes \& 
Wood (1990) and hold as long as $P \la 420$ days. Due to the large number 
of stars, the error bars of the barycenter may be neglected. 

We have also derived from Wood \& Sebo (1996) mean $K_0$ magnitudes and
$(J-K)_0$  colours of O-rich LPVs found near a few clusters of the LMC (2 data
points  per star, taken at arbitrary phases). The magnitudes were then
converted into  bolometric ones by applying the empirical bolometric
correction  ${\rm BC}_K\,=\,f((J-K)_0)$ given by Bessell \& Wood (1984). The
obtained  precision, including intrinsic variability effects, should be roughly
0.3 mag  for $<K>$ and $\le 0.10$ for $<J-K>$. Two thirds of these stars are
obviously  pulsating on a higher-order mode than the Mira-like population,
since they  form a second, parallel strip in the PL plane. That is why we
included this  sample among the LMC data. 

Last, $K$ magnitudes and periods of hundreds of pseudoperiodic red variables 
belonging to the MACHO sample have been found in Wood (1999). Being 
single-phase observations, these data represent the mean magnitude within 
about 0.3--0.5 mag. On the other hand, the periods of these stars are 
secure, since they were derived from MACHO light curves spanning years.

\section{Modelling}
\subsection{LNA pulsation models} 

The AGB linear nonadiabatic (LNA) pulsation models used in this study 
are based on the code of Tuchman et al. (1978), modified as explained in 
Barth\`es \& Tuchman (1994). The equation of state (EOS) includes the 
radiation and an ideal gas of e$^-$, H$_2$, H$_2^+$, H, H$^-$, H$^+$, He, 
He$^+$ and He$^{++}$, as well as a few heavy elements, while abundances are 
determined by solving the Saha equation. 
Convection is treated according to the mixing-length formalism of 
Cox \& Giuli (1968), with instantaneous adjustment to pulsation. 
Recent opacity tables, assuming solar composition and including molecules at 
low temperature, namely OPAL92 and Alexander (1992) [see Alexander \& 
Ferguson 1994], are used. 

The grid of models assumes $X=0.7$ and covers metallicities $Z=0.02$ and 
0.001, masses ranging from 0.8 to 2 $M_\odot$, luminosities ranging from 1000 
to possibly 50000 $L_\odot$, and three values of the mixing--length 
parameter\,: $\alpha=1$, 1.5 and 2. Then, a log-linear least-squares fit 
(excluding the extreme luminosities where isomass lines would turn back) 
gives us theoretical relations between the effective temperature or pulsation 
mode periods on the one hand, and the mass, metallicity, luminosity and 
mixing-length parameter on the other (MLZ$\alpha$T and MLZ$\alpha$P 
relations). Between $\alpha=2$ and 3, the extrapolated periods and effective 
temperatures are precise within 1--2 \%.\\ 

\begin{figure*}[t] 
\vspace{7cm}
\includegraphics{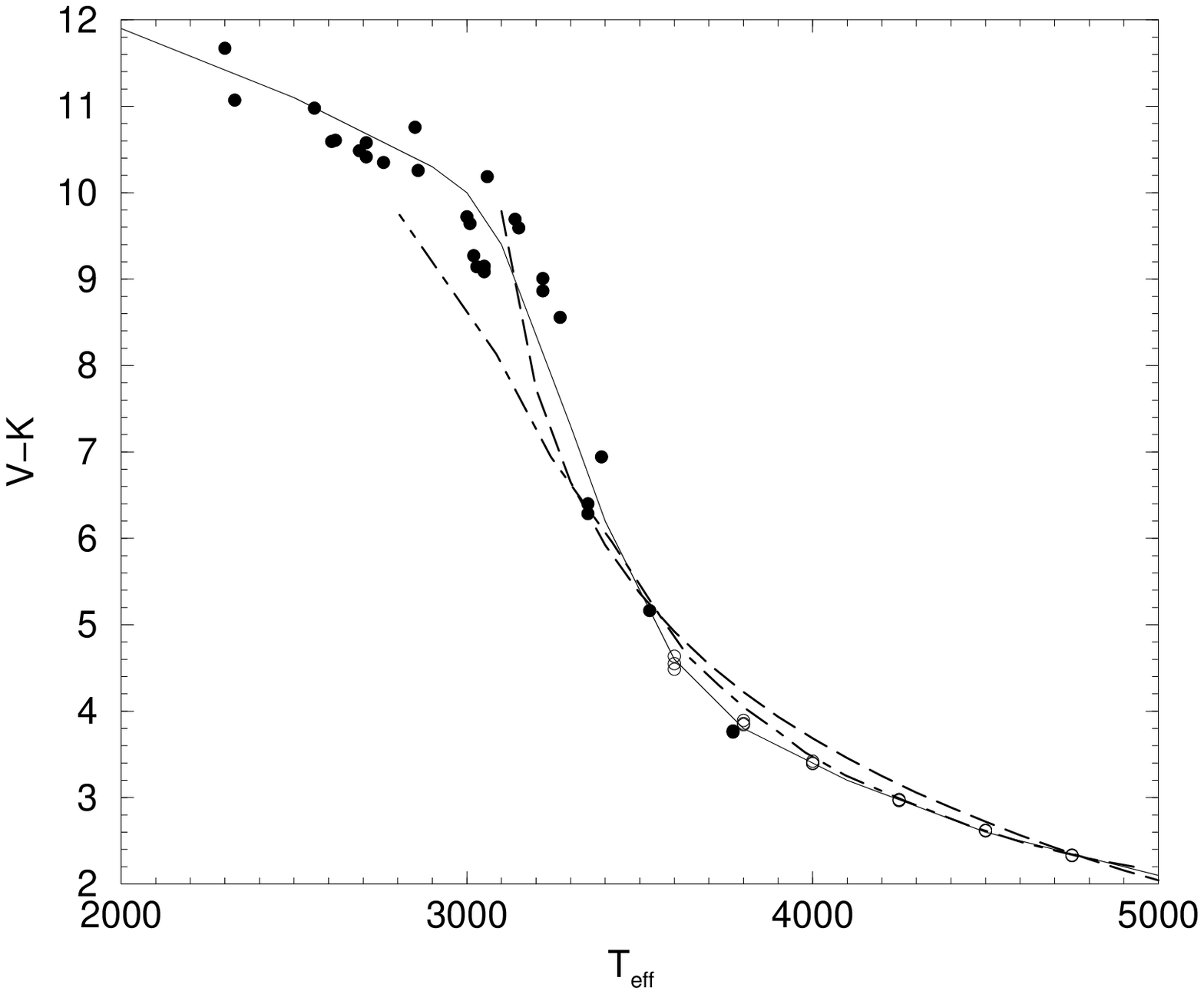}
\includegraphics{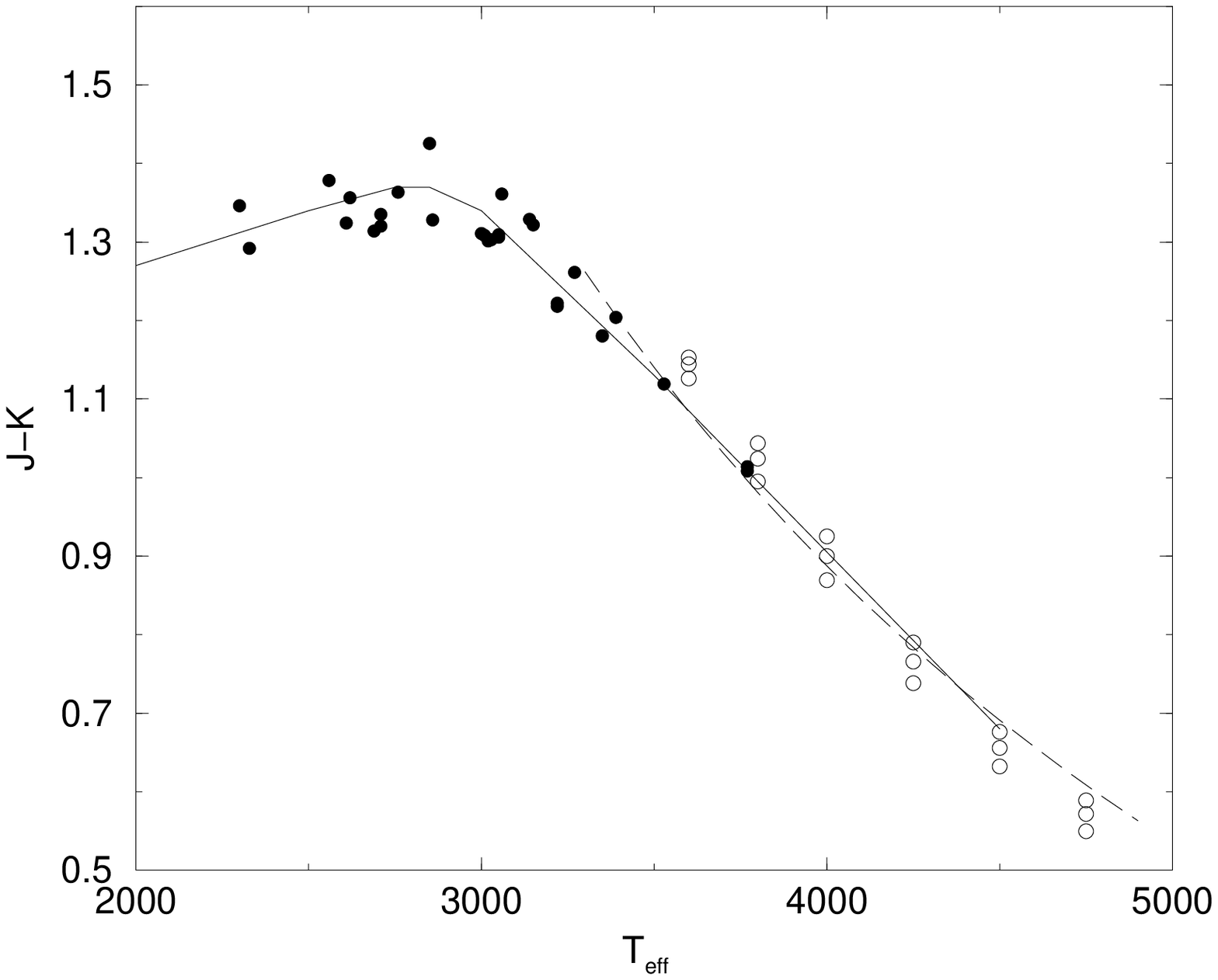} 
\caption[]{{\bf a.} Adopted relation between $V-K$ and effective temperature 
(full line), compared to dynamical models of Bessell et al. [1989a, 1996] 
(filled circles), static models of Bessell et al. [1998] (open circles) and 
the empirical non-LPV relations of van Belle et al. [1999] (dashed line) and 
Perrin et al. [1998] (dot-dashed line). Solar metallicity assumed.\\ 
{\bf b.} The same for $J-K$. The empirical non-LPV relation (dashed line) 
is from Bessell et al. (1983).} 
\label{fig1}
\end{figure*} 

\subsection{Temperature scale and bolometric correction} 

Comparison between the theoretical models and observational data requires 
one to convert effective temperatures into $V-K$ or $J-K$ indices. 

Bessell et al. (1989a, 1996) have computed dynamical models 
of Mira atmospheres with various fundamental parameters (but always 
{\it solar} metallicity) and derived many colour indices, including 
$V-K$ and $J-K$, at various phases, thus various effective temperatures 
(up to 3770 K), gravities and atmospheric extensions. Recently, Bessell et al. 
(1998) have published broad-band colours derived from {\it static} models of 
giant star atmospheres for temperatures higher than 3600 K. Comparison with 
the older static models of Bessell et al. (1989b) (which were the basis of 
their dynamical models) shows that the recent ones are systematically redder 
by a few $10^{-1}$ mag for $V-K$ and a few $10^{-2}$ mag for $J-K$ in the 
small overlapping domain of temperature and gravity. This is due to  
differences in the opacities and their handling. 

On the other hand, van Belle et al. (1999) have published an empirical 
relation between $V-K$ and $T_{\rm eff}$ (above 3030 K) derived from the 
interferometric angular diameters of 59 {\it non-LPV} giant stars 
(non-variable or with a very small $V$ amplitude). 
Similar work, extending to lower temperatures, was also performed by Perrin 
et al. (1998). Moreover, an empirical relation involving $J-K$ has been 
published by Bessell et al. (1983), again for non-LPV stars. 

Considering the inconsistency of all these sources, there is no better way  to
derive a colour--temperature (CT) relation for each colour index than to
perform a simple eyeball fit to the model data, as shown in Fig.~1.  This
takes into account the above-mentioned dynamical models of Bessell  et al.
(1989a, 1996), and the static models of Bessell et al. (1998) at  gravities
$\log g=-0.5$, 0 and $+0.5$. Roughly, one may estimate that our  empirical
relations are precise within 0.5 mag for $V-K$ and 0.05 for $J-K$  (subject to
possible systematic error due to the imperfect modelling of the  molecular
lines).\\ 

Of course, it was necessary to correct the CT relations for intrinsic 
metallicity-dependence. The only information that we could find in the 
literature originated from static models. So, at temperatures $\leq 3350$ K, 
we applied a parabolic \{$\Delta(\log Z)$, $\Delta$(colour)\} fit and a 
series of linear \{$T_{\rm eff}$, $\Delta$(colour)\} interpolations to the 
models XX, X, Y and YY of Bessell et al. (1989b). They correspond respectively 
to $[M/H]=+0.5$, 0, $-0.5$ and $-1$, and to $\log g$ evolving from $-1.02$ 
to $-0.43$ as the temperature increases, so as to mimic the AGB. 
At $T_{\rm eff} \geq 3600$ K, we proceeded in the same way with the models 
listed in Table 5 of Bessell et al. (1998), while extrapolating the gravity 
sequence initiated by the models of Bessell et al. (1989b). 

Around 3480 K, the \{$T_{\rm eff}$, $Z$, $V-K$\} relation resulting from 
Bessell et al. (1989b) exhibits a crossing-over that appears only around  3800
K in the Bessell et al. (1998) models. As a consequence, the $V-K$  correction
was simply interpolated between 3350 and 3600 K, without  considering the
intermediate model data. This does not concern $J-K$.\\ 

On the other hand, wherever necessary (see Sect. 6), we have converted the 
theoretical bolometric magnitudes into $K$ magnitudes, by subtracting the 
the empirical bolometric correction ${\rm BC}_K\,=\,f(V-K)$ given by 
Bessell \& Wood (1984). The used $V-K$ is, of course, the value derived 
from the model temperature.

\section{Is theory directly comparable to observations ?} 

Comparing theoretical models with observational data, with a view to 
calibrating the internal parameters of the theory and to deriving physical 
information on the stars (e.g. the pulsation mode and the fundamental 
parameters) does not make sense if the models are basically wrong or if  their
predictions are systematically shifted because of imperfections of  their input
physics. We thus have to assess the reliability of the  model grid, including
the temperature scale, and to find a way to (partially)  compensate its
defects. 

\subsection{Static/LNA modelling} 

First, one may remember that the core mass--luminosity relation (CMLR)  assumed
in our calculations corresponds to maximum hydrogen shell luminosity 
(Paczy\'nski 1970). In fact, even neglecting the very short helium flashes, 
the luminosity varies by a factor of two over a thermal pulse cycle, and the 
CMLR approximately holds over only about 25\,\% of the interflash time 
(Boothroyd \& Sackman 1988a, 1988b ; Wagenhuber \& Tuchman 1996). For a  global
investigation of our sample of stars, we would perhaps do better adopt  an
"effective" CMLR that would be, say, 30\,\% less luminous. Doing so, at  given
total mass and luminosity, the effective temperature becomes 1\,\% higher,  and
the period decreases by a few \%. 

Our calculations also assume that the convective flux and velocity instantly 
adjust themselves to pulsation, i.e. that the mean eddy lifetime is about 
zero. In fact, it is roughly a third of the fundamental period (Ostlie \&
Cox 1986), which induces a significant phase lag. In order to estimate the 
resulting uncertainty, we have recomputed some of our models while assuming 
a frozen-in convection, i.e. infinite eddy lifetime. The obtained fundamental 
periods are shorter by 5--10\%, and the first overtone longer by 1--4\%. 
In other terms, as long as convection phase lag is concerned, the fundamental 
periods predicted by our model grid are probably overestimated by roughly 
5\,\%, while the first overtone is underestimated by perhaps 2\,\%. 

On the other hand, Ostlie \& Cox (1986) have attempted to improve the standard 
LNA modelling by a horizontal opacity averaging scheme, which accounts in a 
simplified way for the coexistence of rising and falling convective elements 
in the same mass shell. Periods then increase by less than 10\,\%.  Moreover,
the same authors have investigated the effects of  turbulent pressure\,: the
obtained period shifts range from $+8$ to $+36$\,\%  for the fundamental mode
and from $+3$ to $+11$\,\% for the first overtone.  Turbulent viscosity and
energy have negligible effects (Cox \& Ostlie  1993). 

Summarizing, one may expect the model grid described in Sect. 3 to 
underestimate the fundamental period by as much as 40\% for the fundamental 
mode and 25\% for the first overtone, and the relative shift of the latter 
mode is always more than a third of the former.\\ 

It is worth noting that the often quoted models of Wood (Wood 1974, Fox \& 
Wood 1982, Wood 1990, Bessell et al. 1996, Hofmann et al. 1998, Wood et al. 
1999), which include a phase-lagged convection scheme, unfortunately use an 
outdated equation of state. As far as we know, this is the only important difference 
with Tuchman's code. The EOS is thus probably the reason why Wood's linear 
fundamental and first overtone periods are longer by 15--70\% than the ones 
predicted by our models, with the same opacity tables and composition  
regardless of our treatment of convection\footnote{Wood's periods increase
much  faster than ours with luminosity and metallicity}. The relative shift is 
always about the same for the two modes. Interestingly, the fundamental  period
shift is the same order of magnitude as that wich would result from  opacity
averaging and turbulent pressure together. As a consequence, Wood's 
fundamental pulsation models can often reach a reasonable agreement with the 
observations, but first overtone periods derived from this code are strongly 
overestimated, as also noted by Xiong et al. (1998).

\begin{figure*}[!t] 
\vspace{7cm}
\includegraphics{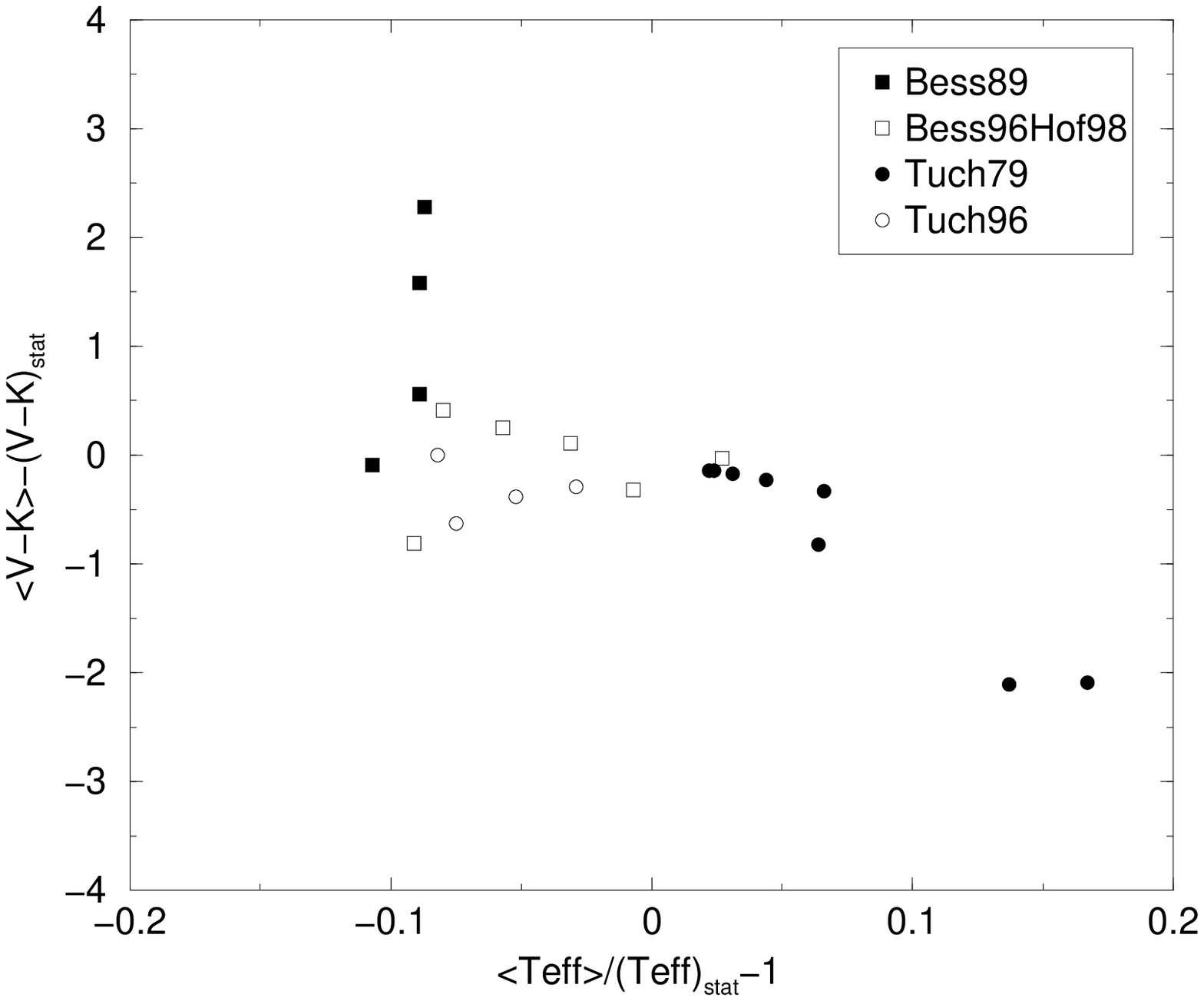}
\includegraphics{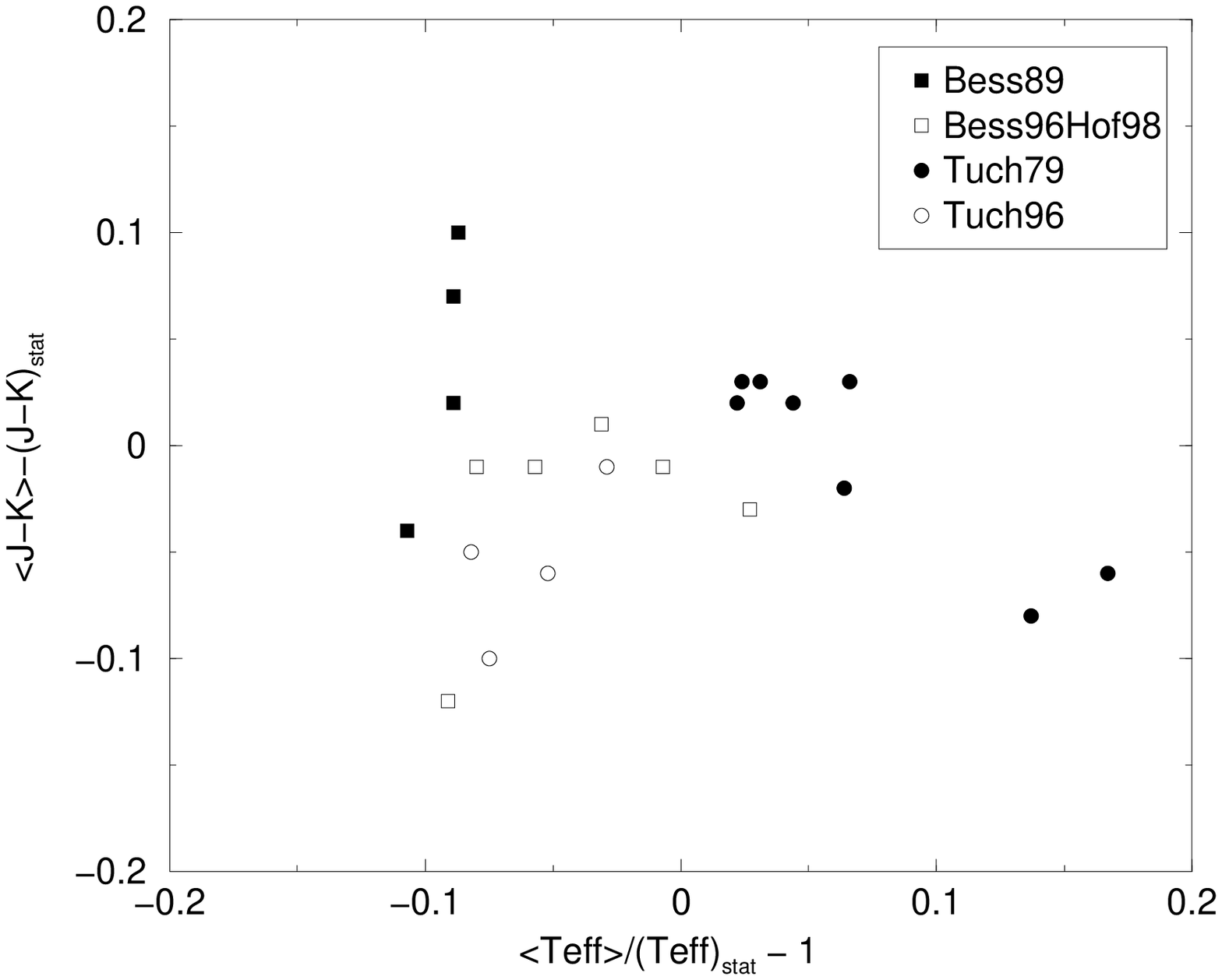} 
\caption[]{Difference between the mean colour index and the colour at the 
static temperature, according to various dynamical models (see text). 
{\it Left:} $V-K$\,; {\it Right:} $J-K$.} 
\label{fig2}
\end{figure*}

\subsection{Nonlinear effects} 

The periods of the pulsation modes predicted by LNA models correspond to 
small-amplitude oscillations of the static stellar envelope. The effective 
temperature, too, is that of the static star. In fact, the  large-amplitude
pulsation of an LPV is strongly nonlinear and makes the outer  envelope expand.
As a consequence, the mean value of the effective  temperature, as well as the
periods and growth-rates of the pulsation modes,  do not necessarily equal the
values given by a linear model of the same star. 

According to the various hydrodynamical calculations performed up to now 
(Wood 1974, Tuchman et al. 1979, Perl \& Tuchman 1990, Tuchman 1991, Ya'ari 
\& Tuchman 1996 \& 1999, Bessell et al. 1996) the static effective temperature 
may differ from the mean $T_{\rm eff}$ of the corresponding pulsating star by 
plus or minus 1--5\,\%. 

On the other hand, some recent calculations (Ya'ari \& Tuchman 1996, 1999) 
have shown that the period of the nonlinear fundamental mode may, after 
thermal relaxation, be {\it shorter} than the LNA value by as much as 35\% 
(depending, at least, on the luminosity). But models based on Wood's code 
(Wood 1995, Bessell et al. 1996, Hofmann et al. 1998) predict only a small 
{\it increase} of the fundamental period, even after spontaneous thermal 
relaxation (models P, M and O of Hofmann et al. [1998]). As stated above, the 
main difference between these two families of models are the treatment of 
time-dependent convection and the equation of state, which both have  important 
thermal effects. As phase-lagged convection tends to increase the 
nonadiabaticity of the pulsation, it is likely that the EOS is the main 
cause of the more ''quiet'' behaviour of Wood's models (as long as the numerical 
scheme is not at stake).

\begin{figure}[!t]
\vspace{7cm}
\includegraphics{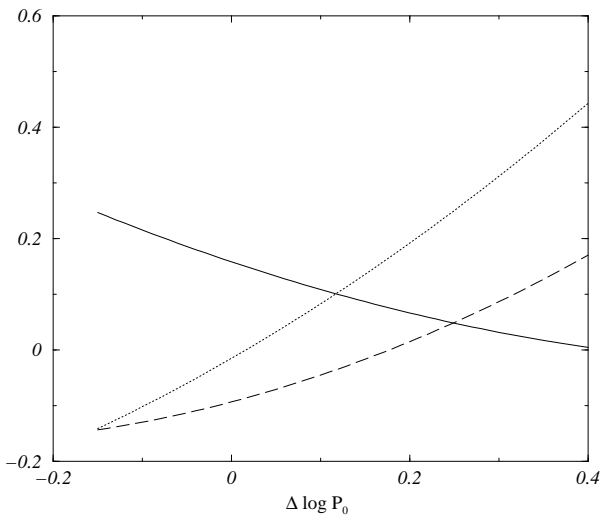}
\caption[]{The $\Delta(J-K)$ colour correction parameter (dashed line), the 
mixing-length parameter $\alpha/2-1$ (dotted line) and the mass discrepancy 
of the fundamental mode in SMC-like clusters $\delta M$ (solid line) as a 
function of the period correction parameter $\Delta\log P_0$.} 
\label{fig3}
\end{figure}

\subsection{Outer boundary} 

Ya'ari \& Tuchman (1996) report that their nonlinear results appear 
basically insensitive to the various outer boundary conditions that they 
have tried. However, all abovementioned models neglect the fact that dust 
condensates in the levitating circumstellar layers and, as an effect of 
radiation pressure, generates a significant stellar wind, ranging from 
$10^{-8}$ to $10^{-6} M_\odot/$yr for Miras and semi-regulars (Jura 1986, 
Jura 1988, Jura et al. 1993, Kerschbaum \& Hron 1992). The physics of this 
phenomenon was extensively described by Fleischer et al. (1992) and H\"ofner 
\& Dorfi (1997). 

Due to the extension of the envelope, the outgoing waves are only partially 
reflected in the photospheric region, and this does not occur at a fixed level 
but at the sonic point, which depends on the wind. Pijpers (1993) has shown, 
in the case of polytropic AGB star models, that the adiabatic fundamental period 
may increase by more than a factor 5 if a mass-loss rate of $10^{-6} M_\odot$/yr 
is assumed. It would be surprising if such a large shift of the adiabatic periods 
had no effect on the nonadiabatic ones. Even though polytropic models are just a 
rough aproximation, one may also expect some effects of this kind in real stars.

\subsection{Temperature scale} 

The nonlinear shape of the temperature--colour relations is another source 
of difficulty\,: it may generate a significant mismatch between the mean 
colour index and the colour corresponding to the mean temperature, depending 
on the latter and on the pulsation amplitude. 

It is thus clear that a linear model having the fundamental parameters 
of a given LPV (in particular its static temperature), the correct mixing 
length, and predicting the correct period, will nevertheless disagree with 
the observed colours by a significant amount. In order to estimate the overall 
colour mismatch ($<V-K>-(V-K)_{\rm stat}$, $<J-K>-(J-K)_{\rm stat}$) that 
may be expected we have applied the above-defined temperature--colour 
relations to simulated temperature oscillations based on the dynamical models 
of Tuchman et al. (1979), Bessell et al. (1989), Bessell et al. (1996), 
Ya'ari \& Tuchman (1996) and Hofman et al. (1998), which have very diverse 
fundamental parameters and amplitudes (usually 1 $M_\odot$, but up to 6 
$M_\odot$ in Tuchman et al. [1979]\,; 1800--35000 $L_\odot$\,; 2250--3500 K 
[static]\,; 15--40 ${\rm km\,s}^{-1}$). The results are shown in Fig.~2. 
The mean shifts and standard deviations are $-0.15$ and 0.95 for $V-K$ 
(but 4/5 of the models lie within 0.7 of the mean), and $-0.01$ and 0.05 for 
$J-K$. {\it No} obvious correlation with any fundamental parameter or with the 
pulsation period or amplitude was found. It is likely that the colour 
mismatch depends on a non-trivial combination of these parameters.

\subsection{Consequences} 

All this discussion leads us to conclude that no linear model grid can 
be expected to directly fit the observations, and that the existing 
nonlinear models are not yet able to provide a reliable grid, or even 
an {\it a priori} correction of the linear models. 
As a consequence, linear models must be complemented by 
additional free parameters, to be added prior to 
comparison with the observations. As a first-order approximation, these 
parameters can be some systematic corrections of the colour ($\Delta(V-K)$ or 
$\Delta(J-K)$) and, for each pulsation mode, of the period ($\Delta\log P$). 
Before trying to interpret the results of Paper~I, these parameters and 
the mixing length have to be calibrated by fitting the models to independent 
data, namely the LPVs observed in globular clusters and in the LMC.

\section{Models calibration\,: clusters and LMC stars} 

The sample of globular cluster stars can be divided in three sets\,: one with 
the metallicity of the LMC ($Z=0.008$)\,; a second one with about the 
metallicity of the SMC ($Z=0.004\pm0.001$)\,; and the last one with 
$Z\approx 0.0005$. 
Only the two first sets (hereafter called LMC-like and SMC-like) will be 
used, because the adopted log-linear model fit and colour--temperature 
relation are no longer reliable at very low metallicities and masses. In 
fact, each set may be divided in two sub-sets, obviously corresponding to two 
different pulsation modes (they are well separated in the PL and PC planes, 
see Fig. 5). Each set will be represented by two points, namely the 
barycenters (mean periods, magnitudes and colours) of its two 
sub-sets\footnote{The calculation of the means excludes one SMC-like star 
lying right in between the two subsets, and three outlying LMC-like stars}. 
This facilitates the model fitting, reduces the observational error bars 
down to low levels ($\le 0.03$ on $J-K$), and reduces the possible effects 
of the scattering of $<J-K>-(J-K)_{\rm stat}$ down to $\sigma \la 0.015$. 
In fact, there are only four stars in LMC-like globular clusters. We thus 
prefered to merge them with the sets of LMC stars (i.e. two with the 
Mira-like stars and two with the higher-order pulsators).\\ 

As an additional constraint, we assume that the mean mass of the Mira-like 
population of the LMC is 1 $M_\odot$. This ensures that, whatever the choice 
of the free parameters, the masses of the sample LPVs of the LMC do 
not exceed 1.5 $M_\odot$, consistent with the evolutionary calculations 
(Wood \& Sebo 1996). Then, for every value of $\Delta\log P$, the theoretical 
MLZ$\alpha$T and MLZ$\alpha$P relations give us the single possible value 
of $\alpha$ and of $\Delta(J-K)$. The pulsation mode that gives the better 
fit to Mira-like stars is always the fundamental. 

Then, keeping the three parameters unchanged, we obtain two masses for the 
barycenter of the SMC-like subset corresponding to the fundamental mode\,: 
one derived from the period, the other from the colour. The calibration thus 
consists in minimizing the difference $\delta M$ of these two masses. 
As shown in Fig.~3, the mass discrepancy decreases as the period (and colour) 
shift increases. The model fit starts being acceptable ($\delta M \le 0.1 
M_\odot$) at $\Delta\log P_0 = +0.13$. The mean colour shift derived from the 
dynamical models (Fig. 2), viz. $\Delta (J-K) = -0.01$ and thus 
$\Delta\log P_0 =+0.16$ and $\alpha=2.29$, gives a reasonable fit\,: 
$\delta M = 0.07 M_\odot$. 
We adopt this solution, for which the most probable value of $\Delta (V-K)$ 
is already known ($-0.15$, of course, i.e. the mean of the {\it a priori} 
estimates of Sect. 4). Indeed, considering all uncertainties (especially 
concerning $<J-K>$ and the CTZ relation) as well as the lack of 
solid theoretical ground, it would make little sense to look for an exact 
agreement of the masses by further increasing the correction parameters 
(see Sect. 7 for further discussion). 

As a last step, we can now determine the period correction of the first 
overtone, which must ensure that consistent masses are obtained from the 
MLZ$\alpha$T and MLZ$\alpha$P relations at the barycenters of the subsets 
corresponding to this mode. 
We obtain $\Delta\log P_1= 0.056$, with negligible mass discrepancies. 

The so-calibrated model grid is represented in Figs. 4 \& 5 by a series of 
isomass lines (0.6, 0.8, 1 and 1.5 $M_\odot$). The results of this 
calibration are summarized in the upper-left quarter of Table 1. Let us 
however recall that masses as low as 0.6 $M_\odot$ are probably slightly 
underestimated by the log-linear fit scheme.\\ 

The likelihood of the calibrated model grid can be checked by confronting it 
to the $K$ magnitudes and periods of the MACHO sample of Wood (1999). 
As can be seen in Fig. 6, the strip corresponding to Mira-like stars 
still fits the fundamental mode, with mean mass 1 $M_\odot$. The strip 
immediately on its left fits the first overtone. Last, a third strip obviously 
corresponds to the second overtone. We could not calibrate its correction 
parameter but, having noticed that $\Delta \log P_1 \simeq {1\over 3} \Delta 
\log P_0$, we adopted $\Delta \log P_2 = {1\over 3} \Delta \log P_1$. 
Our interpretation is consistent with that of Wood et al. (1999) and 
Wood (1999), who also considered the stars lying on the right of the figure 
as probable binaries or stars pulsating on a thermal mode coupled to the 
fundamental.\\

\begin{table*}[!t] 
\caption[]{Mean theoretical masses and metallicities of the fundamental, first 
overtone and second overtone pulsators of the LMC, Globular Clusters with 
SMC metallicity and the solar neighbourhood. The mixing-length parameter 
is assumed to be constant (left) or to decrease by 35\,\% every -2 bolometric
magnitudes along the AGB (right).} 
\begin{flushleft} 
\begin{center} 
\begin{tabular}{lccccccc} 
\hline 
\noalign{\smallskip}
  & \multicolumn{3}{c}{$\alpha=2.29$} & &\multicolumn{3}{c}{$\alpha=f(L)$}\\ 
\noalign{\smallskip}
\hline 
\noalign{\smallskip}
                   & Mode & $<Z>$ & $<M>$ & & Mode & $<Z>$ & $<M>$ \\  
                   &      &     &$(M_\odot)$& &     &    &$(M_\odot)$  \\ 
\noalign{\smallskip}
\hline 
\noalign{\smallskip}
LMC  (Mira-like)  & F & 0.008 & 1.00 & & F & idem  & idem \\
LMC               & 1ov & 0.008 & 0.95  &  & 1ov & idem  & 0.95  \\
$Z_{\rm SMC}$ GC  & F & 0.004 & 0.8~\,  &  & F & idem  & 0.75 \\
$Z_{\rm SMC}$ GC  & 1ov & 0.004 & 0.6~\,  &  & 1ov & idem  & 0.5~\,  \\ 
Group 1 pop.       & F & 0.020 & 0.9~\, & & F & 0.020 & 0.9~\, \\
\mbox{}\hspace{1cm} sample & F & 0.04~\, & 1.7~\, & & F & 0.024 & 1.85 \\ 
Group 2 pop.        & F & 0.027 & 0.95 &  & F & 0.07~\, & 0.85 \\
\mbox{}\hspace{1cm} sample & F & 0.035 & 1.2~\, & & F & 0.05~\, & 1.1\,\, \\
Group 3 pop.       & 1ov & 0.04~\, & 1.1~\, &  & 1ov & 0.07~\, & 1.0~\, \\
                   & 2ov & 0.04~\, & ~\,0.75$^a$ & & | & | 
                   & | \\
Group 4 pop.        & F & 0.009 & 1.1~\, &  & F & 0.010 & 1.1~\, \\ 
\mbox{}\hspace{1cm} sample & F & 0.014 & 1.65 & & F & 0.010 & 1.7~\, \\ 
\noalign{\smallskip} 
\hline
\noalign{\smallskip}
$\Delta\log P_0$&\multicolumn{3}{l}{$+0.16$}& &\multicolumn{3}{l}{$+0.126$}\\ 
$\Delta\log P_1$&\multicolumn{3}{l}{$+0.056$}& &\multicolumn{3}{l}{$+0.033$}\\
$\Delta\log P_2$&\multicolumn{3}{l}{$+0.019$}& & \multicolumn{3}{l}{$+0.009$}\\
$\Delta(J-K)$&\multicolumn{3}{l}{$-0.01$}& &\multicolumn{3}{l}{$-0.065$}\\
$\Delta(V-K)$&\multicolumn{3}{l}{$-0.15$}& &\multicolumn{3}{l}{$-0.53$}\\ 
\noalign{\smallskip}\hline
\mbox{}$^a$ Lower boundary. 
\end{tabular} 
\end{center} 
\end{flushleft} 
\label{tab1} 
\end{table*}

\section{Application to the solar neighbourhood} 

Having fully calibrated the models, we can now investigate the results  of
Paper~I, i.e. the four de-biased PLC distributions of the local LPVs and  the
calibrated and de-biased individual data (sample stars). For the sake of 
clarity, members of the old disk (Groups 1 \& 2 as defined in Paper~I), 
thin-disk (Group 3) and exended-disk/halo (Group 4) populations will be 
separately investigated. Mathematically speaking, the work to be done  consists
in determining the pulsation mode, mass and metallicity  corresponding to the
barycenter of the de-biased distribution of each group,  by solving the
MLZ$\alpha$T and MLZ$\alpha$P equations with the  above-calibrated free
parameters.\\ 

At each barycenter, the absolute $K$ magnitude was converted into the 
bolometric one by applying the bolometric correction defined in Sect. 3.2., 
using the corresponding $<(V-K)_0>$. Then, the period and colour  correction
parameters ($\Delta\log P=0.16$ or 0.056 and $\Delta(V-K)=-0.15$)  were
subtracted from the data points before solving the two equations.  It must be
mentioned, however, that Figures 7 through 11 were plotted  keeping the
observations unchanged, i.e. using $K$ magnitudes.  We thus applied to the
models bolometric corrections that were deduced from  the theoretical $V-K$
after adding $\Delta(V-K)$. The results are detailed  in the next subsections
and summarized in the left-hand part of Table 1.  The isomass lines in the
figures correspond to 0.8, 1 and 1.5 $M_\odot$.\\ 

The de-biased 3D distribution of each Group, shown in Fig. 7,  appears as a 
quasi-elipsoidal volume containing 60\,\% of the population. In the three 
fundamental planes, it is represented by a quasi-elliptical contour which is 
the projection of the elliptical $2\sigma$ iso-probability contour defined in 
the main symmetry plane (see Paper~I). Then, the pulsation mode and the mean 
mass and metallicity are given by the barycenter of the de-biased 
distribution, i.e. the center of the ``ellipsoid'' or ``ellipse''\,: if the 
models have been well calibrated and if their adopted metallicity equals the 
actual population mean, the surface corresponding to one theoretical mode in 
the 3D diagram must include the barycenter. In the three 2D figures, this 
point must lie on the same iso-mass line, which property was used above for 
calibrating the models. Another advantage of working on the barycenter is  that
we avoid the projection effects that occur when the ``ellipsoid''  crosses the
single-mode single-metallicity PLC surfaces with a high incidence.

\begin{figure*}[htb]
\vspace{7.cm}
\includegraphics{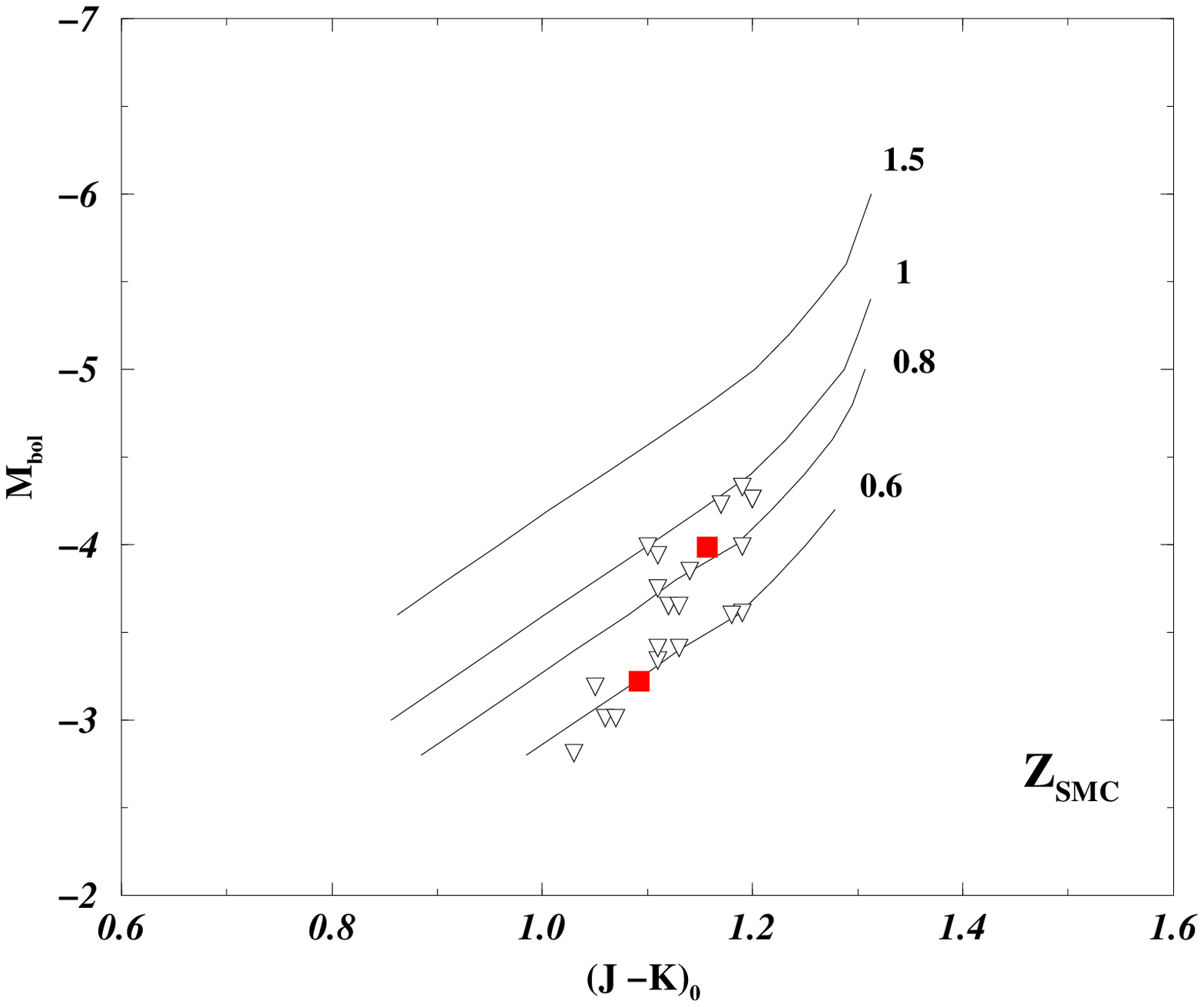} 
\includegraphics{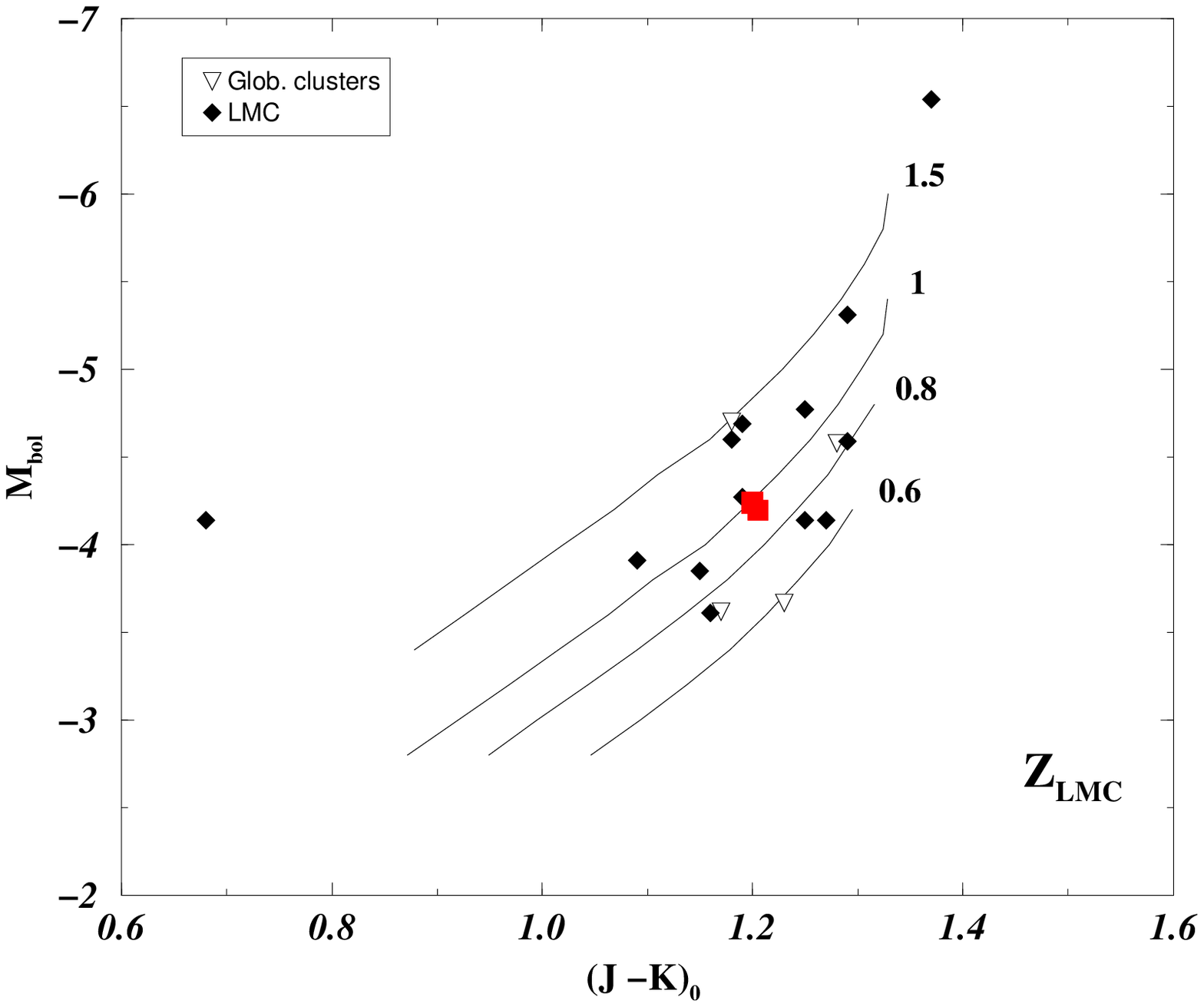} 
\caption[]{{\it Left:} The LC distribution of LPVs in globular clusters with 
SMC-like metallicity compared to the calibrated models, assuming $Z=0.004$. 
The two barycenters are indicated by filled squares.\\ 
{\it Right:} The same for clusters with LMC-like metallicity and for the 
barycenter of the Mira-like population of the LMC with $P \le 420$ d\,; 
$Z=0.008$ assumed.} 
\label{fig4}
\end{figure*}

\begin{figure*}[htb]
\vspace{12.cm} 
\includegraphics{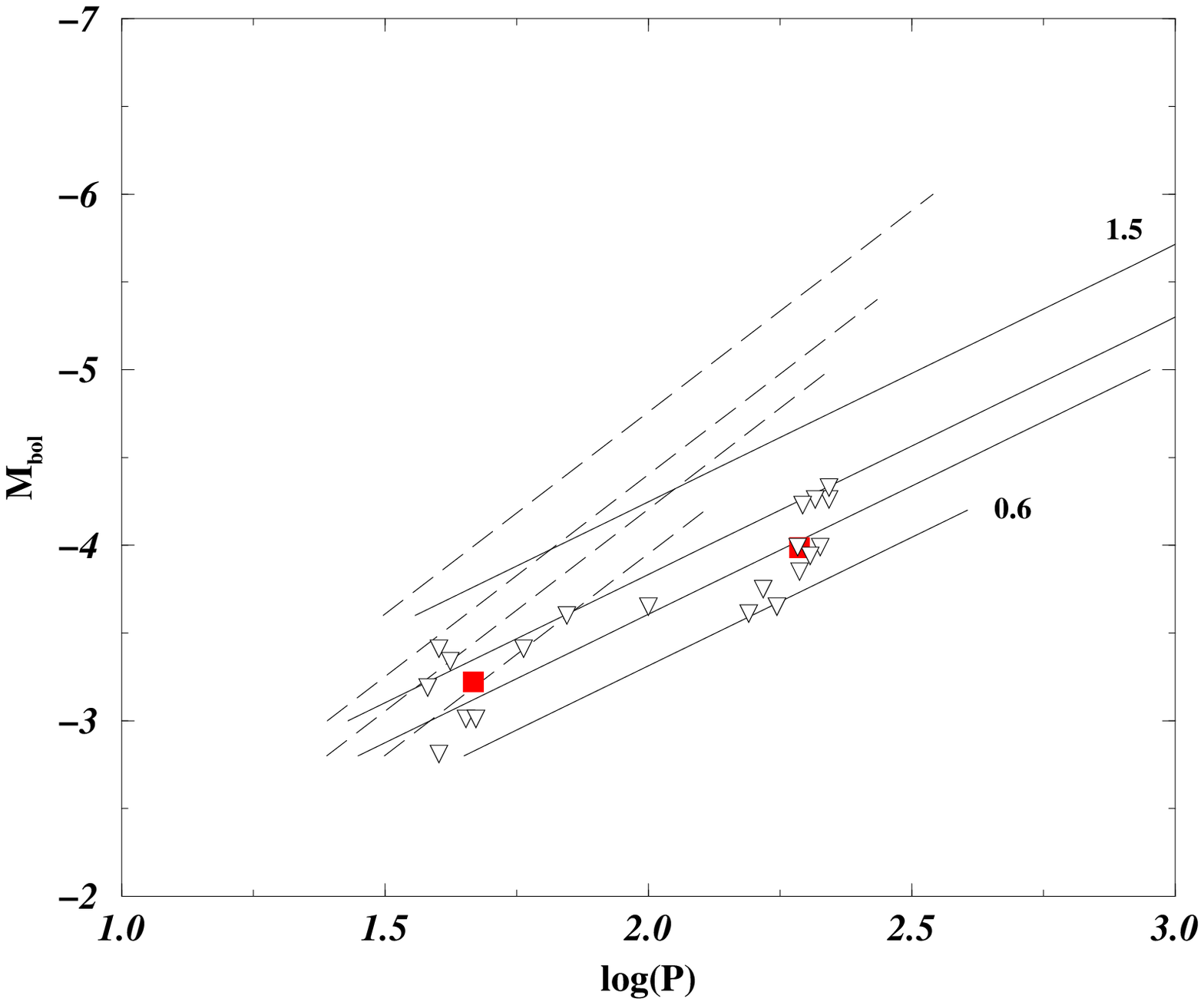} 
\includegraphics{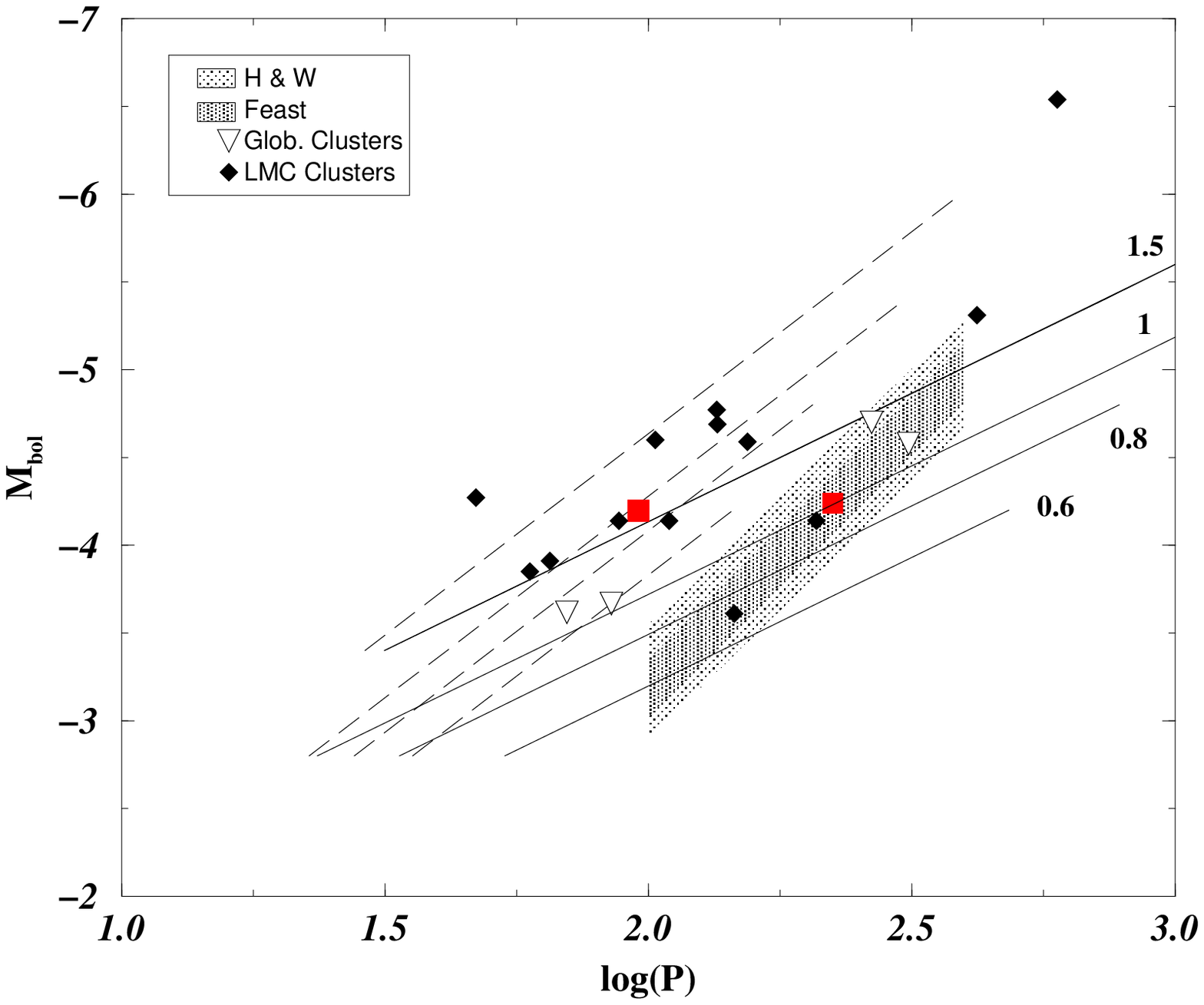} 
\includegraphics{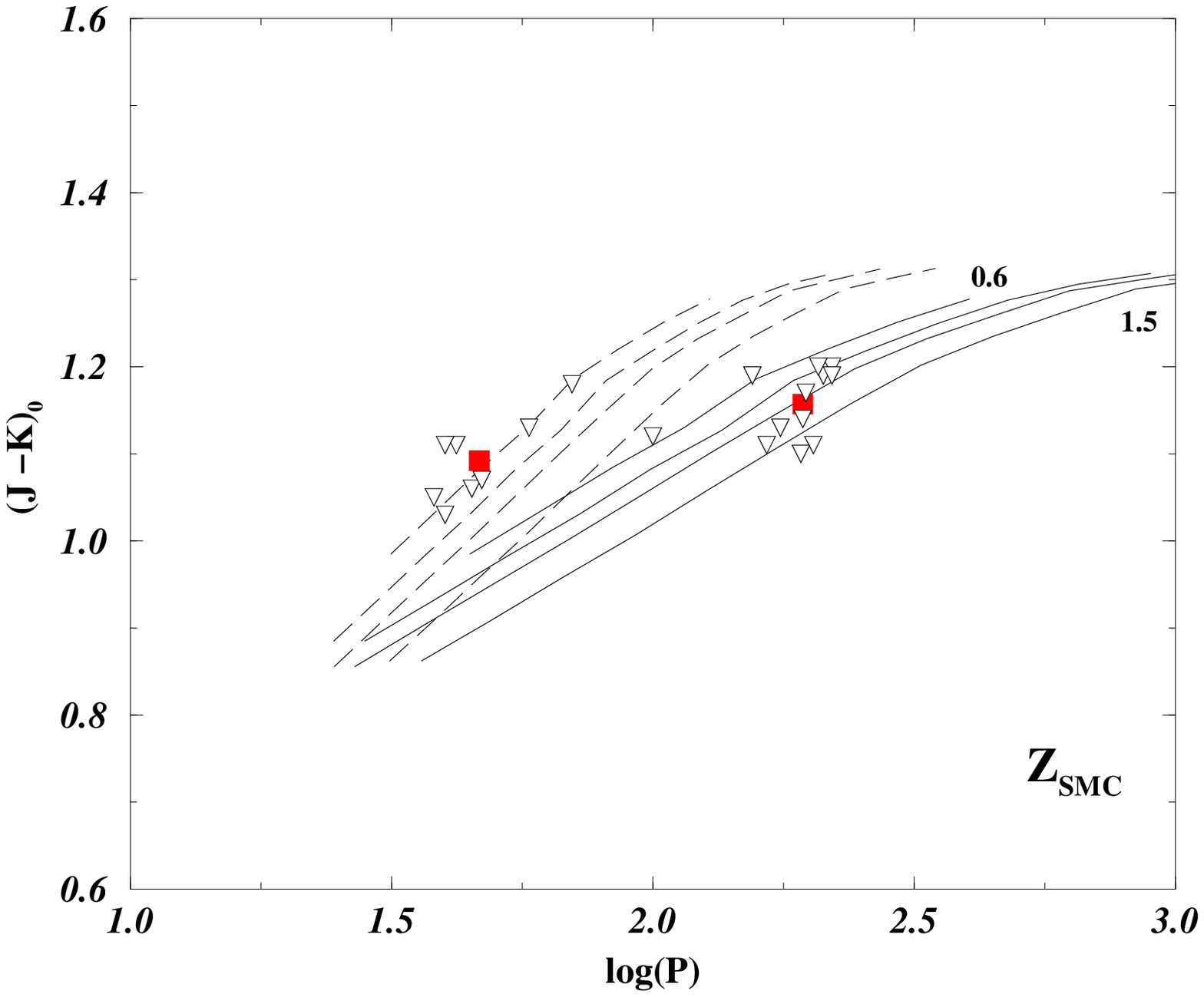} 
\includegraphics{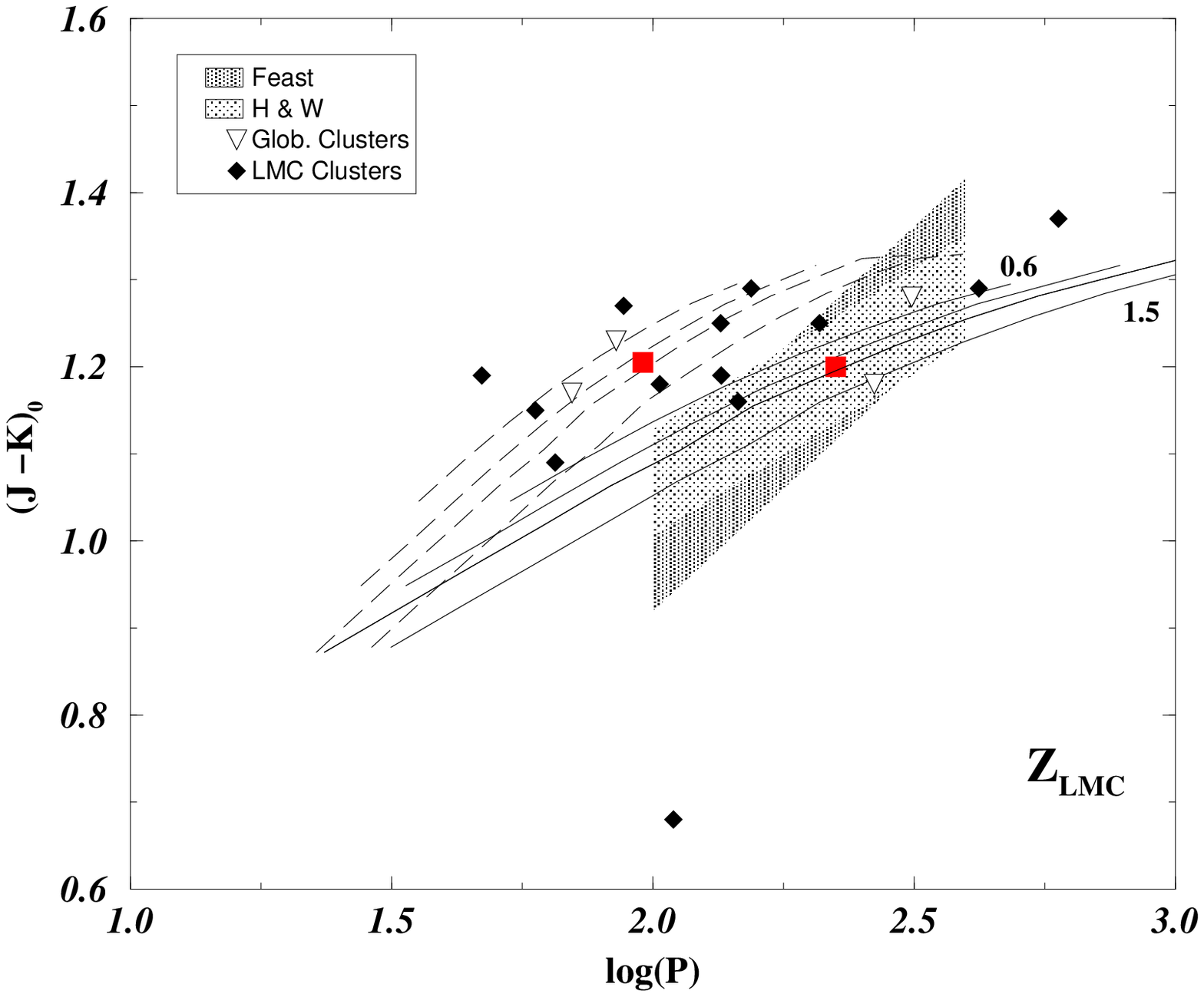} 
\caption[]{The PL ({\it top row}) and PC ({\it bottom row}) distributions of 
LPVs in globular and LMC clusters and the ones of LMC Mira-like stars, 
compared to the calibrated models\,: fundamental mode (solid lines) and first 
overtone (dashed lines); filled squares indicate the single-mode 
barycenters of the data.\\ 
{\it Left:} SMC-like metallicity. {\it Right:} LMC-like metallicity. The 
superimposed linear strips correspond to the whole Mira-like population of the 
LMC with $P \le 420$ d\,: mean PL and PC relations (and the standard 
deviations about them) given by Feast et al. (1989) and Hughes \& Wood 
(1990).} 
\label{fig5}
\end{figure*}

\subsection{Old disk stars} 

For Group 1, the only possible pulsation mode is the fundamental. The 
metallicity is $Z=0.02$ and the mass 0.9 $M_\odot$. On the other hand, 
Group 2 pulsates on the same mode, but with $Z=0.027$ and $M=0.95 M_\odot$. 

As in the preceding section, a careful look at the PL and LC diagrams 
on the one hand and the PC on the other, shows a contradiction if a single 
metallicity is assumed within Group~1. Indeed, when 
increasing the period and colour, the mass decreases in the latter plane, 
while it increases in the two former. This is the 2D translation of the fact 
that, in 3D, the main symmetry plane of the population is much inclined 
with respect to the single-$Z$ single-mode theoretical surfaces. If 
$\Delta\log P$ is supposed to increase by any reasonable amount with the 
period, the problem is only slightly attenuated. If $\Delta(V-K)$ too is 
assumed to depend on the period, then it has to reach about $-3$ at 
the top of the Group~1 sample. This is at the limit of (or exceeds) the 
{\it a priori} estimates, which anyway exhibit no obvious correlation with 
the period (see Sect. 4 and Fig. 2). 
Therefore, complete explanation of the PLC distribution of Group 1  
probably requires the metallicity to significantly increase with the period. 
The mass, too, must strongly increase with the period (even if only the 
colour shift is invoked). Actually, if the correction parameters are kept 
unchanged, a metallicity of 0.04 is found at the barycenter of the Group~1 
{\it sample}, together with a mass of 1.7 $M_\odot$. Concerning the Group 2 
sample, the metallicity (0.035) differs just a little from the  
population mean, and the mass is 1.2 $M_\odot$.

\subsection{Thin-disk stars} 

Group 3 stars cannot pulsate in the fundamental mode, since this would 
require $Z=0.44$. If we assume that they are pulsating on the first overtone, 
the mean mass and metallicity are $1.25 M_\odot$ and $Z=0.08$, which seems 
a bit too metallic for stars of the solar neighbourhood. 

However, the MACHO sample (see Fig. 6) suggests that this group may be a 
mixture of first and second overtone pulsators. That is, the ``ellipse'' 
representing the population would, in fact, be a global fit to two  steeper,
less extended ones. Unfortunately, we were unable to separate  these two
sub-groups (modes) when doing the luminosity calibration, probably  because of
the paucity of the sample in view of the obervational error bars.  We are thus
deprived of any direct estimate of the barycenter of each  overtone population.
Nevertheless, we can use the MACHO sample as a guide  to determine two points,
typical of the first and second overtone  pulsators of the Group 3 population.
To this purpose, we first have to shift  the MACHO sample by 0.1 magnitude
(roughly corresponding to the luminosity  and colour shifts involved by the
metallicity difference between the LMC and  our Galaxy), so that the Group 3
``ellipse'' crosses the two bulges at the  bottom of the PL strips, and the
top-left limit of the Group 3 sample matches  that of the MACHO data. Then the
intersections of the  strips with the de-biased period--luminosity relation
defined by the  ``ellipse'' can be picked out. They are found around $\log
P=1.33$ and 1.85. The corresponding  colours are given by the PLC relation
found in Paper I.  The models fit these points with $Z=0.04$ and $1.1 M_\odot$
for the first  overtone pulsators, and the same metallicity but $0.75 M_\odot$
for the  second overtone. The latter value is, in fact, a lower boundary of the
mean  mass, since the adopted point lies at the bottom edge of the second
overtone  population.

\subsection{Extended disk and halo stars} 

Group 4 stars (Figs 8 \& 11) appear pulsating on the fundamental mode with 
$Z=0.009$ and 1.1 $M_\odot$ (0.014 and 1.65 $M_\odot$ for the sample). 
No solution is found for the first overtone. However, these results must 
be taken with some caution, because of the paucity of the sample in view of 
the data error bars (see Sect. 2.1).

\section{Discussion} 

\subsection{A problem of mixing length ?} 

A single value of the mixing-length parameter has been used throughout this
work.  We did not possess  enough data on LMC and globular cluster stars to
securely calibrate  variations of $\alpha$ and of the correction parameters,
but  some discussion of this deserves attention and is given below. 

Comparison between the mixing-length theory and 2D numerical simulations of 
stellar convection indicate that, at the bottom of the RGB, 
the mixing-length parameter should exceed the solar value by at least 5\,\% 
-- or even 10\% according to some 3D calculations. Low-mass K sub-giant 
models have $\alpha=\alpha_\odot+10$--15\,\%, and this parameter is a 
decreasing function of the effective temperature while it increases with 
the surface gravity (Freytag \& Salaris 1999, Ludwig et al. 1999). 

Empirical results are also available\,: Chieffi \& Straniero (1989), 
Castellani et al. (1991), Bergbusch \& Vandenberg (1997), Vandenberg et al. 
(2000) were able to fit the Red Giant Branch (RGB) of Globular Clusters 
(masses $\le 1 M_\odot$) with the same mixing length parameter as for 
solar-like stars. However, Stothers \& Chin (1995) as well as Keller (1999), 
showed that the mixing length parameter must exceed $\alpha_\odot$ by roughly 
35\,\% at 3 $M_\odot$ and strongly decrease at larger masses (and 
luminosities) if the red giants and supergiants of Galactic open 
clusters and of young clusters of the Magellanic Clouds are to be fitted. 
This non-monotonic behaviour illustrates the competition between the mass and 
the luminosity in determining the gravity and temperature. 
No clear metallicity-dependence was found (Stothers \& Chin 1996, Keller 
1999). Concerning the Horizontal Branch and the Early-AGB, Castellani et al. 
(1991) found the solar mixing length to be suitable for Globular Cluster 
stars with masses $\le 0.8 M_\odot$. 

Although the LPVs are TP-AGB stars, with much higher and diverse 
luminosities, these previous works give us the order of magnitude of the 
possible variations of $\alpha$ between the mean mass or luminosity that were 
the basis of the calibration (barycenter of the LMC Mira-like stars) and 
3 $M_\odot$ or the maximum luminosity of the sample. We have thus derived 
a mass--mixing-length relation from the RGB results, by performing a 
spline interpolation and scaling so as to match the $\alpha$ and mass that 
were obtained at the barycenter of the LMC Mira-like stars. 
On the other hand, when investigating the possibility of a 
luminosity-dependence, we have assumed that $\alpha$ varies by $35$\,\% 
every 2 magnitudes.\\ 

If $\alpha$ depends on the mass, the isomass lines in this paper must move 
away from each other in the PL and LC diagrams below $3\,M_\odot$ and get 
closer again at larger masses. This means that, in the previous sections, the 
masses that appeared larger than 1 $M_\odot$ and the corresponding 
metallicities were respectively over- and underestimated. As a consequence, a 
totally unlikely metallicity ($Z > 0.10$) is now required in a large part of 
Group 1. The Group 4 sample, too, gets unlikely high mass and metallicity in 
view of its kinematics. Moreover, the period corrections also have to be 
increased, which makes the model fit more unlikely in view of the theoretical 
background. The mixing-length parameter is thus probably not or is little 
dependent on the mass. The same arguments also allow us to rule out the 
hypothesis that it increase with the luminosity.\\ 

On the other hand, let us assume that $\alpha$ is a decreasing function of 
$L$. Then, the slopes of the isomass lines get smaller in the PL and LC 
diagrams. The same combination of $\Delta \log P_0$, $\Delta (J-K)$ and 
$\alpha_{\rm LMCmira}$ as in the preceding sections gives a better fit, with 
a mass discrepancy $\delta M=0.04 M_\odot$ for the SMC-like GC fundamental 
pulsators. However, we get $\delta M=0.08 M_\odot$ for the first overtone 
pulsators. 
As a compromise, we may adopt $\Delta\log P_0=+0.126$, $\Delta(J-K)=-0.03$ and 
$\alpha_{\rm LMCmira}=2.22$, yielding $\Delta\log P_1=0.033$ and 
$\delta M=0.07 M_\odot$ for both modes. Then, the deviation of $\Delta(J-K)$ 
from the mean of its {\it a priori} estimates (see Sect. 4.4), expressed in  
units of standard deviation, defines a scaling to apply to $\Delta(V-K)$, 
which yields $-0.53$. 

As a first consequence, this hypothesis helps to solve a puzzling 
contradiction that we found in Figs. 4 and 5\,: while the theoretical 
mass increases along the Mira-like strip of the LMC in the PL and PC planes 
it decreases in the PC diagram if a constant mixing length is assumed. 
A bit of metallicity dispersion and of period-dependence of the correction 
parameters might also help. 

The barycenter masses and metallicities of the local, LMC and GC populations 
are shown in Table 1, together with the ones obtained with a constant 
$\alpha$. 
One can see that $<M>$ and $<Z>$ are nearly constant for Groups 1 and 4. 
For Groups 2 and 3 (1st ov.), the mean masses decrease by 10\%, but the 
metallicities reach high values that are unlikely for old stars of the solar 
neighbourhood. 
Moreover, no solution is found for the Group 3 second-overtone pulsators, 
even when varying $\Delta\log P_2$ by a factor 3. All of this, together with 
the very low mass found for GC 1st overtone pulsators, suggests that 
$\alpha$ is actually less luminosity-dependent than the adopted 
rate\footnote{New calculations performed while allowing 
$\Delta\log P$ to vary with the luminosity have shown that $\alpha$ can not 
decrease by more than 8\,\% per magnitude along the AGB if the mean mass 
of the 1st overtone pulsators is required to be $\ge 0.5 M_\odot$.}. 

If we further increase the luminosity-dependence of the mixing  length, smaller
or even null period corrections of the fundamental and first  overtone modes
can be adopted. For example, if $\alpha$ varies by 70\,\%  every 2 magnitudes,
the compromise is reached at $\Delta\log P_0=0.063$,  $\Delta(J-K)=-0.065$ and
$\alpha_{\rm LMCmira}=2.09$, yielding  $\Delta\log P_1=-0.004$. However, the
fit is not better  ($\delta M=0.07 M_\odot$ for both modes) and the mean mass
of the SMC-like GC  first-overtone pulsators becomes definitely unlikely ($0.4
M_\odot$).\\ 

Concluding, the mixing-length parameter probably decreases along the 
AGB, but its variation should not exceed 15\,\% per magnitude. Clearly 
{\it positive} period corrections are anyway required, especially for the 
fundamental mode ($>30$\,\%) but also for the first overtone ($>8$\,\%) and 
the latter relative shift is always smaller than a third of the former. 

\begin{figure}[!h]
\vspace{7cm}
\includegraphics{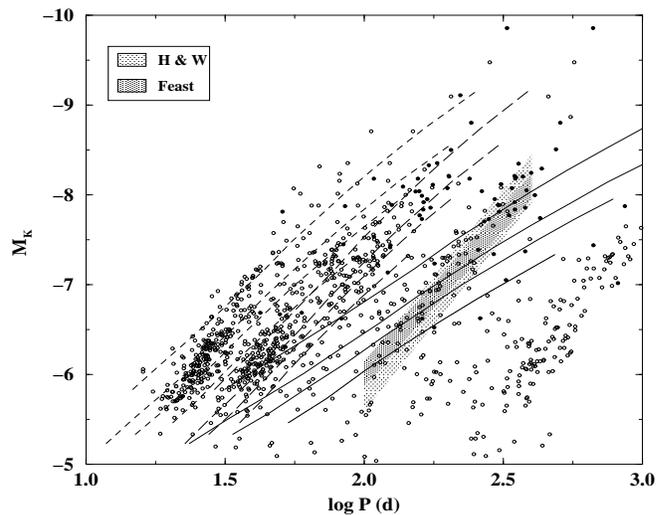} 
\caption[]{PL distribution of red variable stars of the LMC in the MACHO 
data base (magnitudes taken from Wood [1999]), compared to the calibrated 
models\,: fundamental mode (solid lines), first (long-dashed) and second 
(dashed) overtones. Also shown are the Mira-like PL relations of Feast et al. 
(1989) and Hughes \& Wood (1990).} 
\label{fig6} 
\end{figure} 

\begin{figure*}[!p]
\vspace{12cm} 
\includegraphics{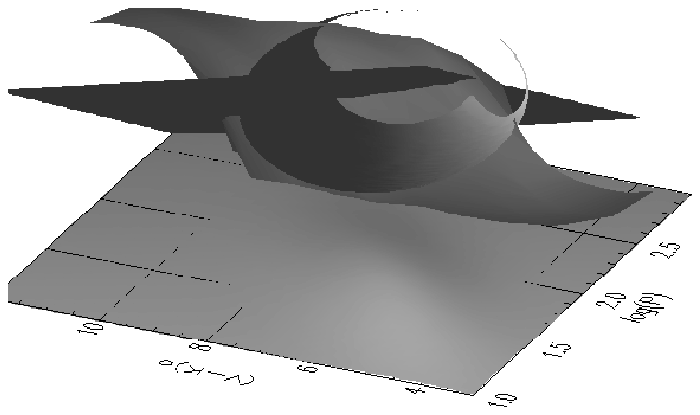}
\includegraphics{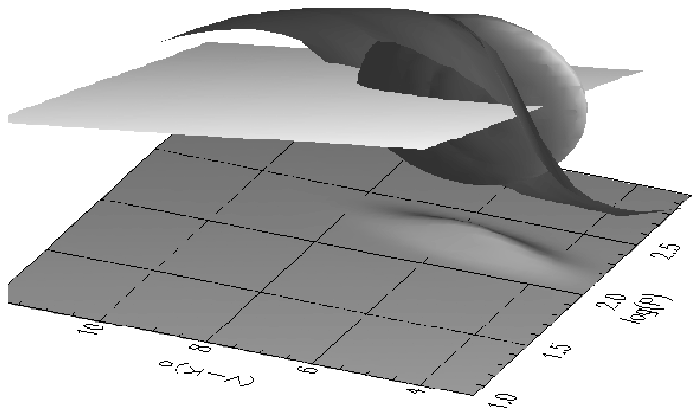}
\includegraphics{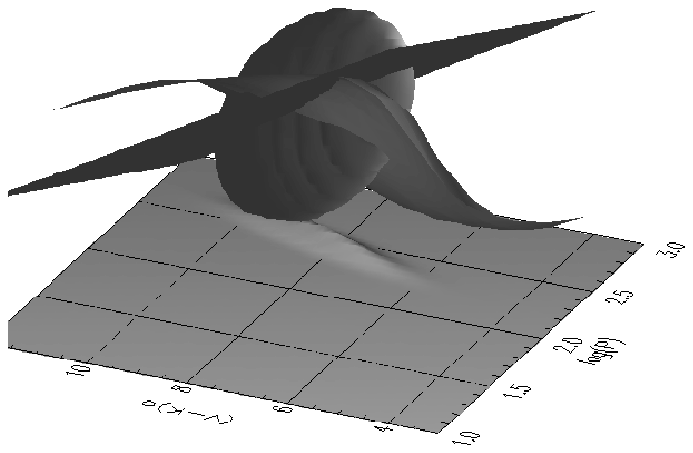}
\includegraphics{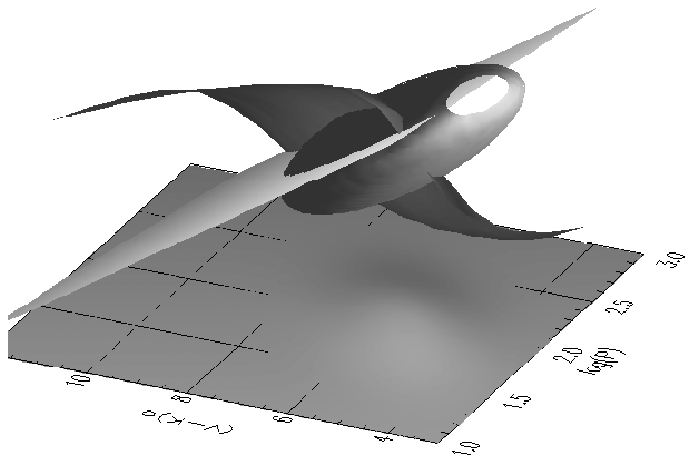}
\caption[]{The de-biased PLC distribution of LPVs in the solar neighbourhood 
({\it top row:} Groups 1 \& 2\,; {\it bottom row:} 3 \& 4), compared to 
calibrated models ($Z=0.02$, 0.02, 0.04 and 0.01 respectively). Each group 
appears as a quasi-ellipsoid containing 60\% of the population. The PLC 
relation (main symmetry plane) and the PC distribution (relief on the $xy$ 
plane) are also represented. The curved surfaces represent the models 
(fundamental mode for Groups 1, 2 \& 4\,; first overtone for Group 3).} 
\label{fig7}
\end{figure*}

\begin{figure*}[!p]
\vspace{7cm}
\includegraphics{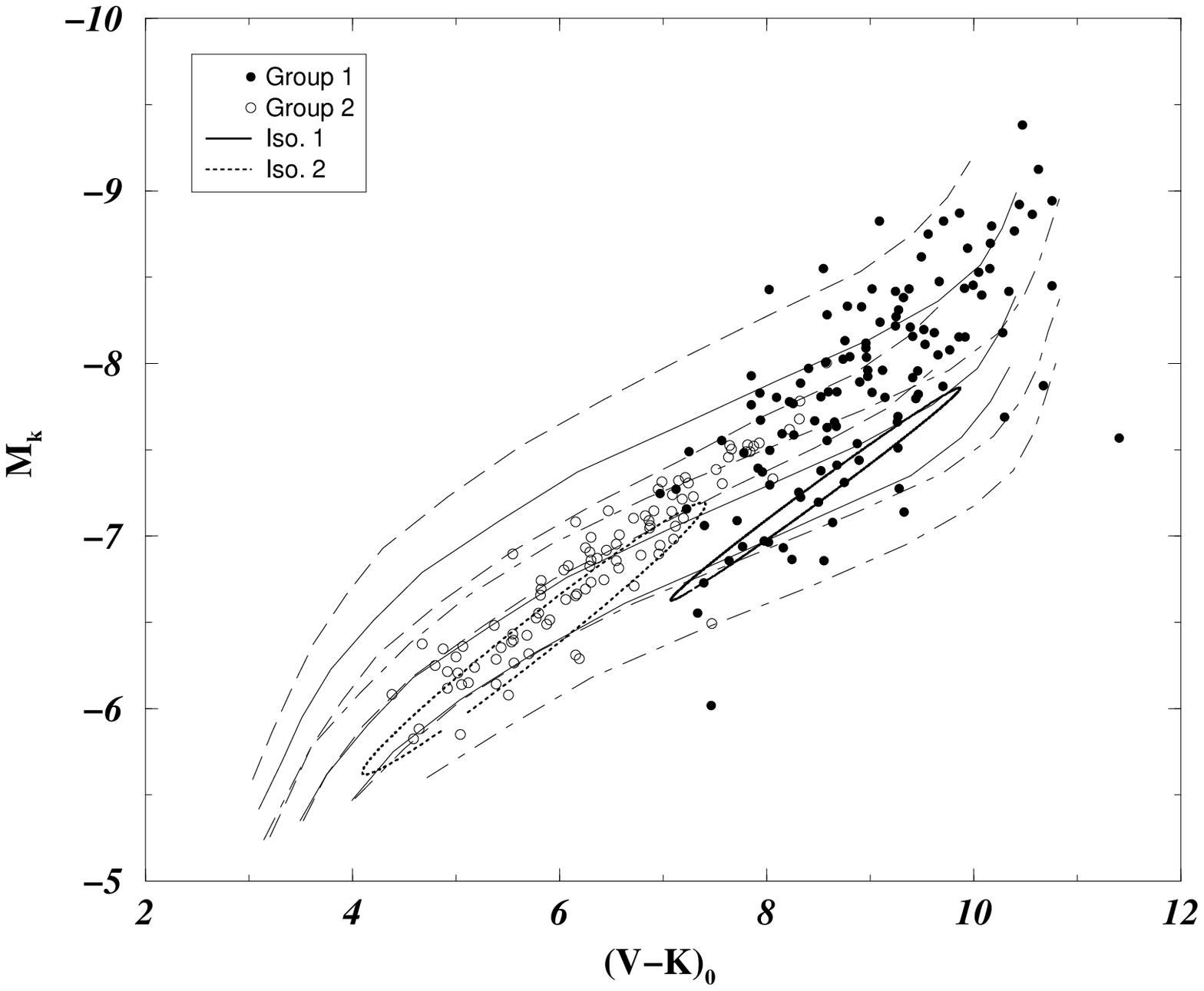}
\includegraphics{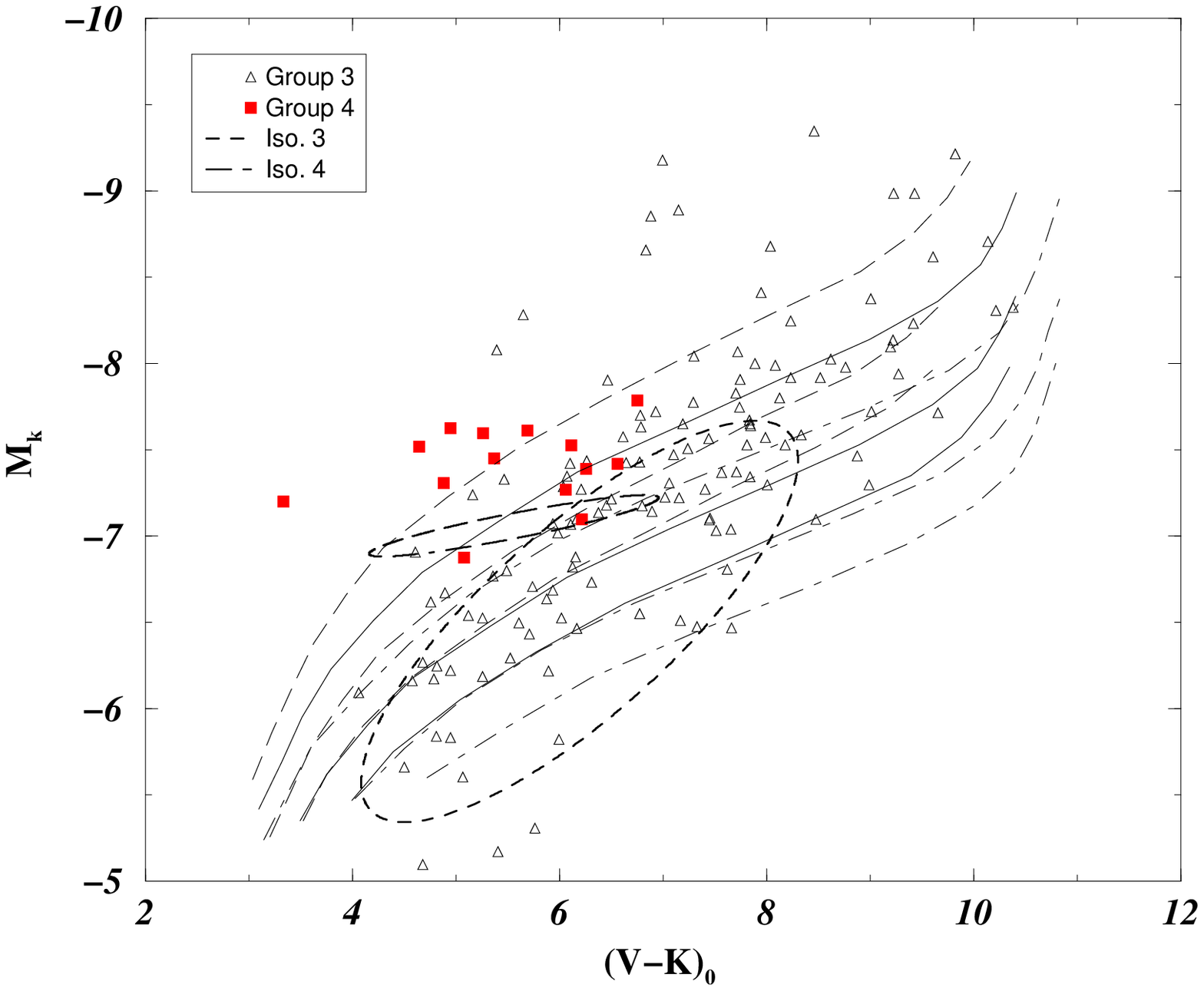}
\caption[]{The LC distribution of local LPVs (sample points and projected, 
de-biased $2\sigma$ iso-probability contours of the populations), compared to 
calibrated models (solid lines\,: $Z=0.02$\,; dashed\,: $Z=0.01$\,; 
dot-dashed\,: $Z=0.04$\,; masses are 0.8, 1 and 1.5 $M_\odot$ from the 
bottom to the top).} 
\label{fig8}
\end{figure*}

\subsection{Stability, consistency and error bars} 

Our results form a consistent set including seven or eight populations with  
five or six different mean metallicities, three pulsation modes and two 
colour indices. If we vary by 0.1 $M_\odot$ the assumed mean mass of the LMC 
Mira-like stars, all other masses simply get shifted by a similar amount. 
The metallicities of the local stars remain unchanged, and the period and 
colour correction parameters vary by only about 0.05 and 0.015 ($J-K$) 
respectively. 

In the single-$\alpha$ case, the period and colour correction parameters were 
taken {\it a minima}, i.e. a better fit (smaller $\delta M$) would have  been
obtained with larger corrections. As a test, let us increase  $\Delta(J-K)$ and
$\Delta(V-K)$ by one standard deviation of the {\it a  priori} estimates (Fig.
2), this corresponds to $\Delta\log P=0.24$ for the  fundamental mode and 0.101
for the first overtone. Then, the mean masses of  Groups 1 through 3 increase
by less than 2\% and the metallicities by only  0.002. This is of course
negligible. Thus, our results are  stable, which is confirmed by their
similarity when a luminosity-dependent  mixing length is adopted (see Table
1). 

On the other hand, the error bars of the periods, magnitudes and colours at 
the barycenter of each population (LMC, Globular Clusters and solar 
neighbourhood) are small (see Sect. 2 and 5) and yield uncertainties of a few 
percent on masses and, for the solar neighbourhood, 5--15\,\% on 
metallicities (except perhaps for Group 4). 

Since any change of the adopted temperature scale would automatically 
translate into the colour correction parameters (the calibrated 
values as well as the {\it a priori} estimates shown in Fig.~2) and into the 
luminosity dependence of $\alpha$, the uncertainty of this relation 
does not significantly affect our results at constant $Z$. However, 
metallicity differences with respect to the LMC might be a little 
overestimated, if molecular opacities were significantly underestimated in 
the atmospheric models.\\ 

Support for this model may be seen in the consistency of the mean masses and 
metallicities of the local populations of LPVs with their respective 
kinematics, determined in Paper~I, and their similar evolutionary stages\,: 

Group~1, which has the kinematics of old disk stars and is mostly 
composed of Miras, is actually found having about the solar metallicity, just 
as usually assumed. 

Concerning Group~3, first and second overtone pulsators have the same, higher 
mean metallicity, consistent with the thin-disk kinematics. 

Also consistently with its kinematics (extended disk and halo), Group 4 has a 
lower metallicity than the others. 

Last, compared to Group 1, Group 2 is found to be much more metallic and a bit
less  massive  ($L$-dependent $\alpha$) or a little more metallic and massive 
(single $\alpha$). Since these stars are slightly less evolved and may  lose
$0.05 M_\odot$ in roughly $10^5$ years, the mean initial mass must be  similar
to or smaller than that of Group 1. So, in both cases, Group 2  stars must be a
little older (Vassiliadis \& Wood 1993). This is consistent  with the larger
velocity dispersion and scale height found in Paper~I. The  evolutionary
aspects of this work will be further investigated in a  forthcoming paper.

\subsection{Theoretical interpretation} 

The significant, positive period correction that has to be applied to 
linear models contradicts the hydrodynamical calculations. Indeed, we have 
seen in Sect. 4.2 that our $\Delta\log P_0$ would range from $-0.19$ to $-0.03$ 
if the models of Ya'ari \& Tuchman (1996, 1998) did represent real stars. This 
would lead to large 
mass discrepancies ($0.1 < \delta M \le 0.35 M_\odot$) for the fundamental 
pulsators of the SMC-like clusters, even with a 
luminosity-dependent mixing length. Moreover, the corresponding 
colour corrections would be extreme values among the {\it a priori} estimates 
calculated in Sect. 4\,: $-0.15\le\Delta (J-K)\le-0.10$ and 
$-2.7\le\Delta(V-K)\le-1.9$. This is quite a bit for an average behaviour\,! 

Another interesting -- though weaker -- argument is derived from 
evolutionary considerations. For the predicted period shifts, the 
mean metallicity and mass ratios of Group 1 to Group 2 are 
$8\,\ge\,{Z_1/Z_2}\,>\,0.8$ and $1.5\,\ge\,{M_1/M_2}\,\ge\,1.1$. 
Detailed evolutionary calculation is beyond the scope of this paper but 
if the periods predicted by Ya'ari \& Tuchman were right, Group 2 would 
probably be a little younger than Group 1, in contradiction with the 
kinematics. 

Summarizing, the nonlinear behaviour calculated by Ya'ari \& Tuchman (1996) 
is unlikely in view of the available data. Let us try to explain this\,:\\ 

As stated in Sect. 4.1, the core mass--luminosity relation assumed in our 
calculations holds over a limited part of the thermal pulsation cycle, but  the
effect on the period and temperature is very small and the  opposite sign of
the calibrated period and colour correction parameters. 

The same section also shows that phase-lagged convection, horizontal  opacity
averaging and turbulent pressure all together might yield a  period increase by
as much as 40 \% for the fundamental and 25 \% for the  first overtone, the
relative shift of the latter always being larger than a  third of the former.
In case of negligible nonlinear effects, this could  explain the correction
parameters of the fundamental or the first overtone  but not both, since the
ratio would be wrong\footnote{The same remark  holds {\it a fortiori} for the
calculations based on Wood's code, which give both modes about the same
shift.}. Furthermore, the nonlinear effects predicted by Ya'ari \& Tuchman
would strongly reduce the final period of the fundamental while not
significantly changing the first overtone, so that both modes would finally
exhibit similar shifts with respect to our LNA models. 

In other terms, if an improved physics of the sub-photospheric regions 
manages to explain the observed first overtone, then there must remain a 
significant, positive shift of the actual fundamental mode ($\ga 15$\,\%, 
possibly much more) with respect to its theoretical period. 
The only explanation seems to be the coupling of the stellar envelope with 
the circumstellar layers and the subsequent wind, evoked in Sect. 4.3 above.

\begin{figure*}[htb]
\vspace{5cm}
\includegraphics{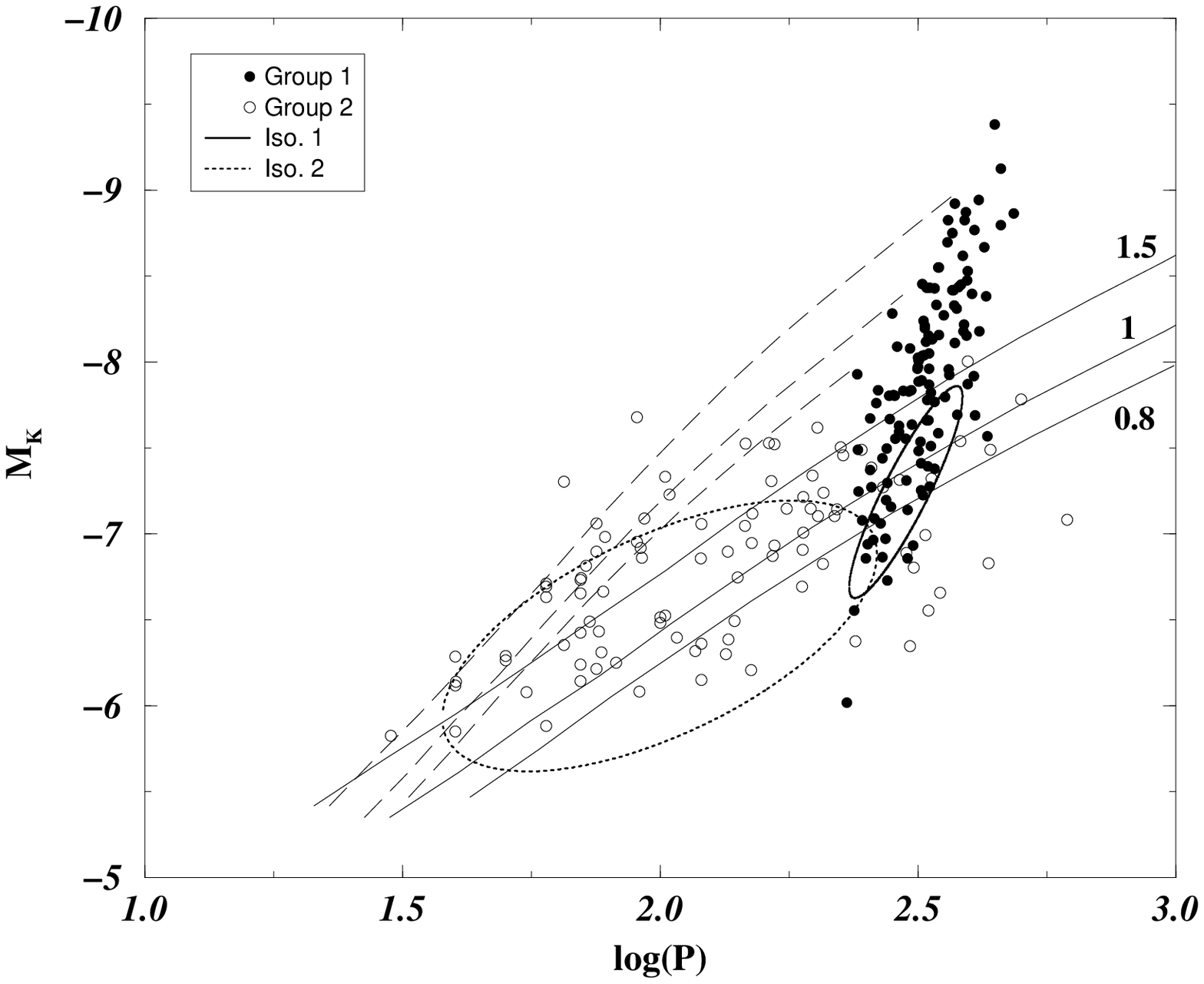}
\includegraphics{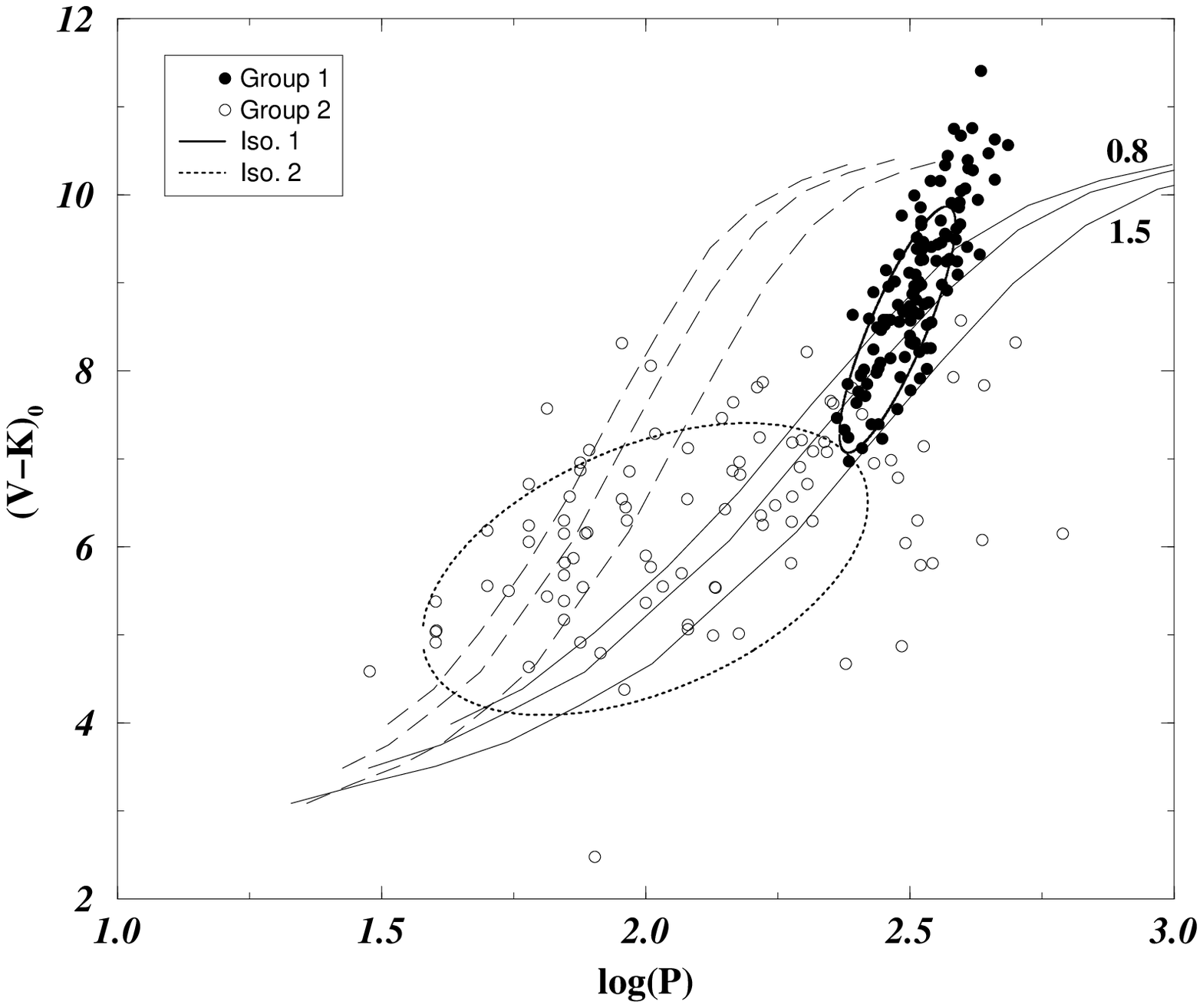}
\vspace{2cm}
\caption[]{The PL ({\bf a}) and PC ({\bf b}) distributions of 
old-disk LPVs, compared to calibrated models with $Z=0.02$: 
fundamental mode (solid lines) and first overtone (dashed lines).} 
\label{fig9}
\end{figure*}

\begin{figure*}[htb]
\vspace{7cm}
\includegraphics{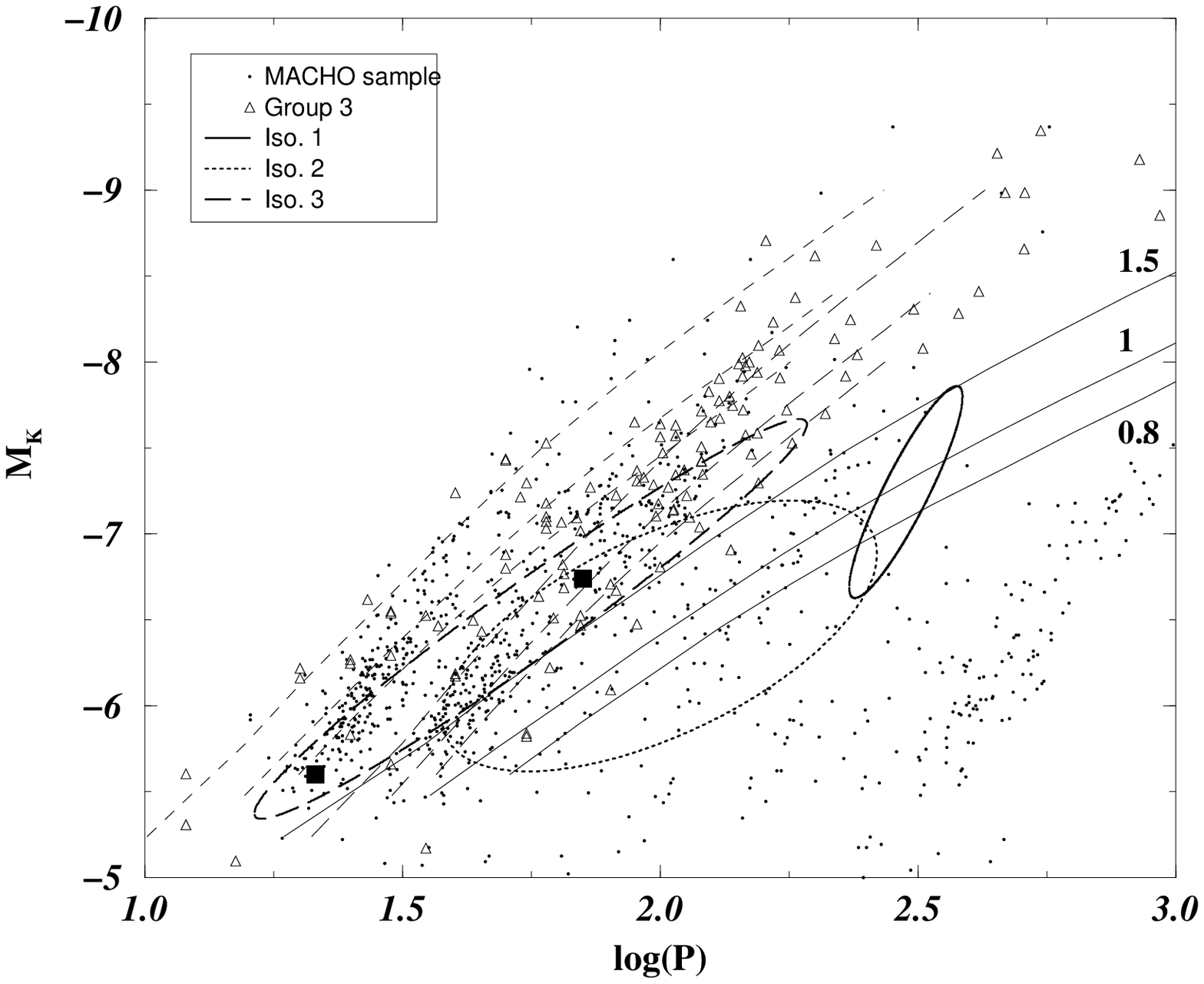}
\includegraphics{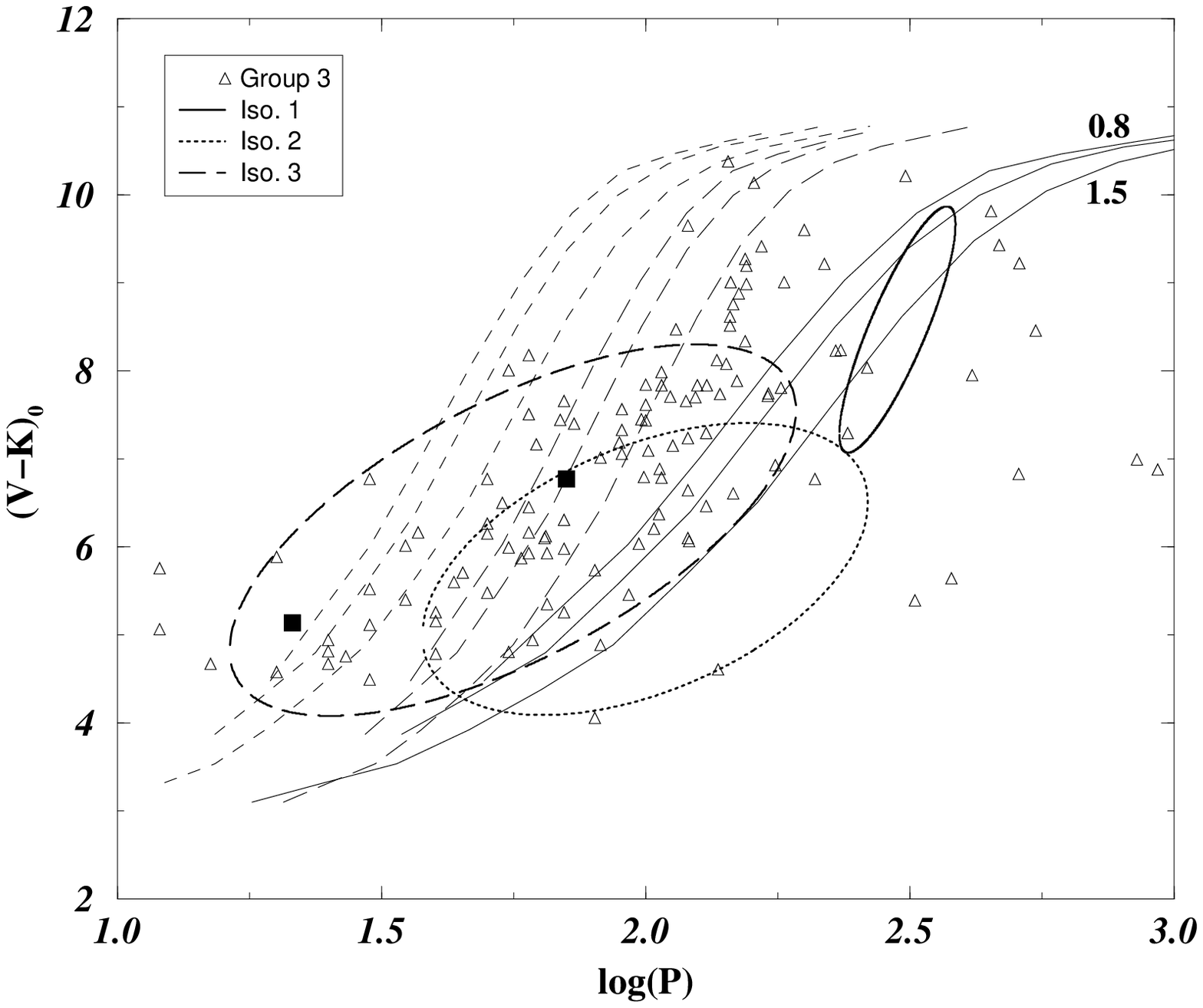} 
\caption[]{The PL ({\bf a}) and PC ({\bf b}) distributions of 
thin-disk LPVs, compared to calibrated models with $Z=0.04$\,: 
fundamental mode (solid lines), first overtone (long-dashed lines) and 
second overtone (dashed lines). Also shown are the de-biased distributions of 
old-disk stars and, shifted by $+0.1$ mag, the MACHO sample (dots). 
The two squares show the adopted typical first and second overtone pulsators 
in the thin-disk population.} 
\label{fig10} 
\end{figure*}

\section{Conclusion} 

The aim of this paper was to interpret the results of Barth\`es et al. (1999)
in terms of pulsation modes and  fundamental parameters,  i.e. the  fact that
the Long-Period Variable stars of the solar neighbourhood are  distributed
among four groups (according to kinematic and photometric criteria),  and to
study the period--luminosity--colour distributions of these groups.  This was
done by confronting them with a grid of linear nonadiabatic pulsation  models. 

Preliminary discussion of the colour--temperature relations and of the 
existing linear and nonlinear modelling codes showed that the periods and 
colours predicted by LNA models may significantly differ from the observed 
values, even if the adopted fundamental parameters are the true ones.  In order
to mimic this behaviour, we added a few free  parameters to the LNA models: a
systematic period correction $\Delta\log P$ for each mode, and  systematic
colour corrections $\Delta(V-K)$ and $\Delta(J-K)$. These  parameters, as well
as the mixing length, were then calibrated by demanding  that consistent masses
be derived from the period and from the colour, for  the fundamental and
first-overtone pulsators of the LMC and of globular  clusters with LMC or SMC
metallicity. It was assumed that the mean mass of  the Mira-like stars
(fundamental pulsators) of the LMC is 1 $M_\odot$. Then,  the mean mass of the
LMC first-overtone pulsators was found to be about 0.95  $M_\odot$, while that
of the fundamental and first-overtone pulsators with  SMC metallicity was 0.8
and 0.6 $M_\odot$ respectively (or down to 0.75 and  0.5 $M_\odot$ if the
mixing-length parameter is allowed to strongly vary with  the luminosity). 

We were thus able to determine the pulsating mode and the mean masses and 
metallicities of the neighbouring LPV populations\,: 

\begin{itemize} 
\item Group~1 stars (old disk LPVs, mostly Miras) are pulsating on the  
fundamental mode. The mean mass is $<M>\,\simeq\,0.9\,M_\odot$ and the mean 
metallicity $<Z>\,\simeq\,0.02$, both strongly increasing with the 
period. 
\item Group~2 stars (old disk LPVs, mostly SRb's) are 
pulsating on the fundamental mode also, with $<M>\,\simeq\,0.9\,M_\odot$ and 
$<Z>\,\ga\,0.03$. Metallicity variations of about 50\,\% are likely. 
\item Group~3 stars (thin-disk LPVs, mostly SRb's) are pulsating on the first 
and second overtones, with \linebreak[4] 
\mbox{$<M_{1ov}>$}$\;\approx\,1.05\,M_\odot$, 
\mbox{$<M_{2ov}>$}$\;>\,0.75\,M_\odot$ and \linebreak[4] $<Z>\;\geq\,0.04$. 
\item Group~4 stars (extended disk and halo) appear to be pulsating on the 
fundamental mode with $<M>\,\simeq\,1.1\,M_\odot$ and $Z\,\simeq\,0.01$. 
This result must however be taken with caution. 
\end{itemize} 

These findings, based on the $V-K$ index as a temperature estimator, have been 
checked and confirmed by using $J-K$. They were proven to be stable and 
consistent with the kinematics. The error bars of the masses are about 5\,\%. 
If one shifts the reference mass of the LMC Miras by any reasonable amount, 
all other masses are simply similarly shifted and the metallicities remain 
nearly unchanged. Metallicity differences with respect to the LMC may be a 
little overestimated if molecular opacities have been significantly 
underestimated in the atmospheric models. Nevertheless, the hierarchy holds 
anyway. 

The mixing-length parameter probably decreases along the AGB, but its 
variation should not exceed 15\,\% per magnitude. 
This was taken into account in the abovementioned results. 

This study confirms the findings in Paper I, the discrimination between Miras 
and semiregulars is not pertinent\,: Groups 1 and 2 not only have similar 
kinematics but also the same pulsation mode.\\ 

It has also been shown that both the linear and nonlinear models that were the 
basis of all previous studies of LPV pulsation are probably far from the real 
pulsational behaviour of these stars. While dynamical calculations including a 
modern equation of state predict a strong reduction of the fundamental 
nonlinear period with respect to the linear one, important, {\it positive} 
systematic corrections have to be applied to the periods of our linear models 
(30--45\,\% for the fundamental mode and 8--13\,\% for the first  overtone).
Improvements of the physics of the sub-photospheric envelope  (phase lagged
convection, turbulent pressure, horizontal opacity  averaging...) appear
insufficient to explain these two shifts altogether, so  that the actual
fundamental period should always exceed the theoretical one  by at least
15\,\%. This led us to conclude that all existing linear and  nonlinear
pulsation codes probably suffer from neglecting  the stellar wind generated by
the interaction with the circumstellar  envelope.\\ 

\begin{figure*}[!t]
\vspace{7cm}
\includegraphics{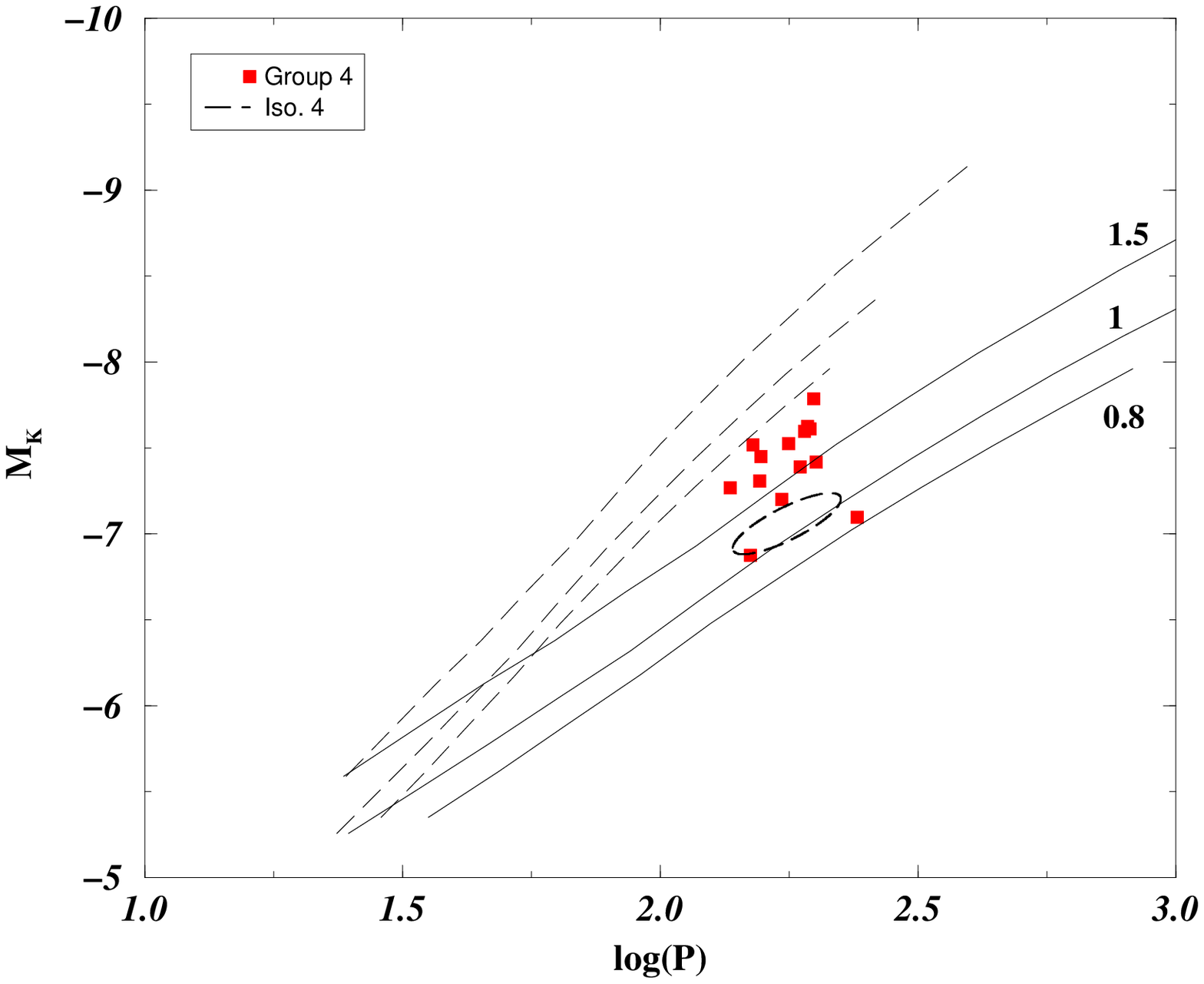}
\includegraphics{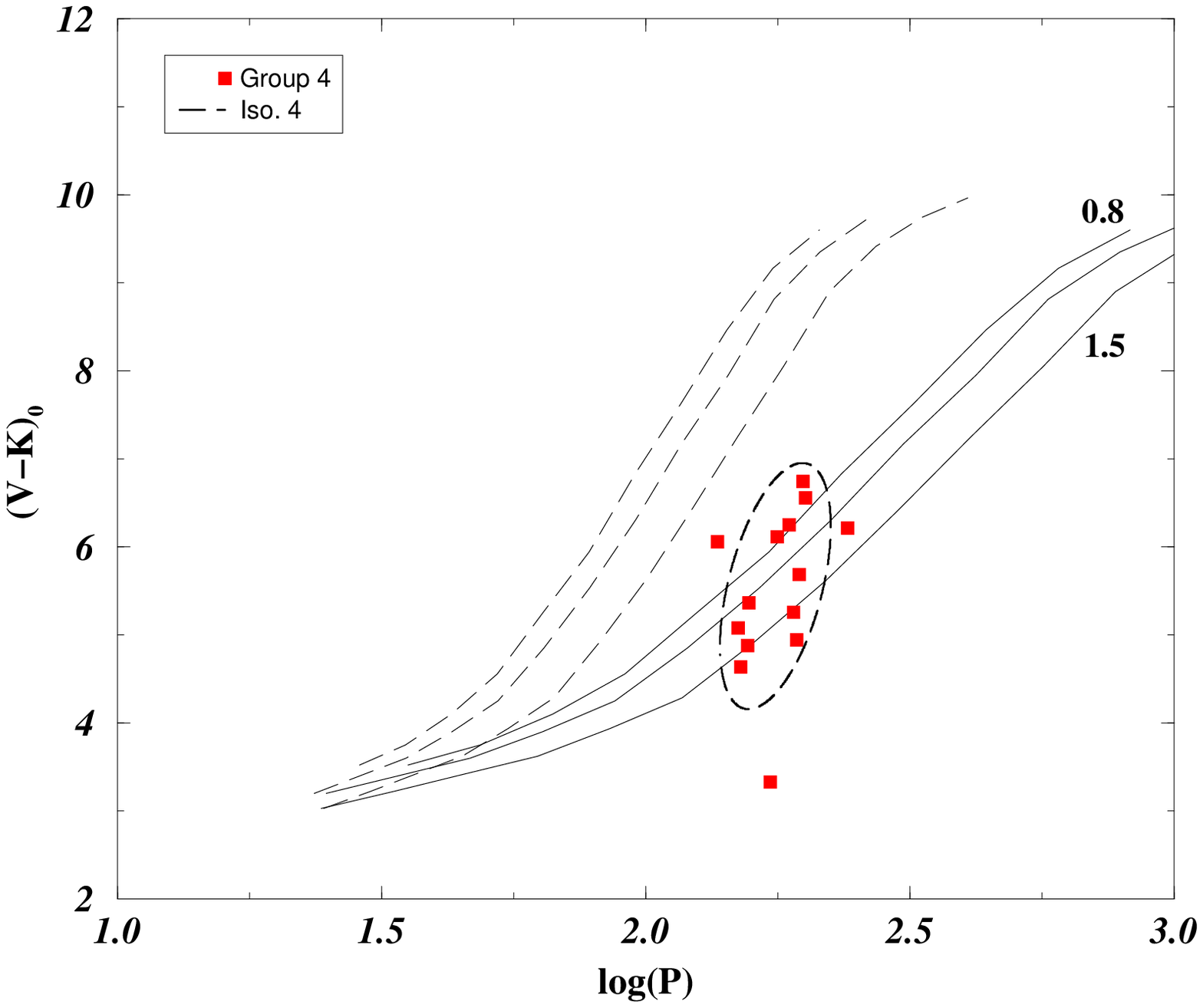}
\caption[]{The PL ({\bf a}) and PC ({\bf b}) distributions of 
extended-disk/halo LPVs, compared to calibrated models with 
$Z=0.01$.} 
\label{fig11}
\end{figure*} 

As a consequence, the works of Barth\`es \& Tuchman (1994) and Barth\`es \& 
Mattei (1997), who confronted LNA models with the Fourier components of the 
lightcurves of a few nearby Miras and concluded in favour of the first 
overtone, should now be reconsidered by taking into account the necessary 
period corrections and the variations of metallicity and, possibly, 
mixing-length parameter. 

This study may also have consequences in AGB evolutionary calculations, 
especially those that use the period as a substitute for a fundamental 
parameter ($M$ or $T_{\rm eff}$) or as the variable in the empirical 
mass-loss function (e.g. Whitelock 1986, Vassiliadis \& Wood 1993, Reid et 
al. 1995, Marigo et al. 1996). Indeed, these studies are based on pulsation 
models by Wood that appear to strongly overestimate the periods and 
their dependence on luminosity and metallicity, probably because of the 
equation of state. Concerning the fundamental mode, this peculiarity allows 
Wood's models to roughly mimic the abovementioned systematic period shifts 
and variation of the mixing length parameter, but with an uncertainty 
that remains to be assessed. Moreover, while many Long Period Variables are 
obviously pulsating on the first or second overtone, this possibility is 
usually neglected in evolutionary calculations, and Wood's models appear 
definitely inapropriate for these modes. These issues will be investigated in 
a forthcoming paper.\\ 

\bigskip

\acknowledgements{This work was supported by the European Space Agency 
(ADM-H/vp/922) and by the Hispano--French Projet International de 
Coop\'eration Scientifique (PICS) N$^{\rm o}$\,348. We thank the referees, 
Drs. Bessell and Bono, for numerous comments that helped us to improve 
this paper.}

\end{document}